\documentclass[11pt,reprint,showpacs,showkeys,nofootinbib,ssfamily]{article}

\usepackage{amsmath,fontenc,calrsfs}
\usepackage{amsfonts,amssymb,mathtools,slashed}
\usepackage{cite} %dash between consecutive cites
\usepackage{hyperref} %manage links
\hypersetup{backref, colorlinks=true  } %remove boxed in hyperref
\usepackage{multirow}
\usepackage[dvips]{graphicx}
\usepackage{float}
\usepackage{appendix}
\usepackage{color}
\usepackage{url}
\usepackage{subfigure}
\usepackage{footnote}
\usepackage{authblk} % template for authors/affiliations

\def\beq{\begin{equation}}
\def\eeq{\end{equation}}
\def\bea{\begin{eqnarray}}
\def\eea{\end{eqnarray}}
\def\nn{\nonumber}

\def\chic1{\chi_{c1}}

\newcommand{\vep}{\varepsilon}

\newcommand{\la}{\langle}
\newcommand{\ra}{\rangle}

\def \Im{\text{Im}\,}

\def\Xint#1{\mathchoice
   {\XXint\displaystyle\textstyle{#1}}%
   {\XXint\textstyle\scriptstyle{#1}}%
   {\XXint\scriptstyle\scriptscriptstyle{#1}}%
   {\XXint\scriptscriptstyle\scriptscriptstyle{#1}}%
   \!\int}
\def\XXint#1#2#3{{\setbox0=\hbox{$#1{#2#3}{\int}$}
    \vcenter{\hbox{$#2#3$}}\kern-.5\wd0}}

\def\dashint{\Xint-}

\newcommand{\vk}{\mathbf{k}}
\newcommand{\vx}{\mathbf{x}}
\newcommand{\vp}{\mathbf{p}}
\newcommand{\vq}{\mathbf{q}}

\usepackage{hyperref}% http://ctan.org/pkg/hyperref
\hypersetup{%
  colorlinks = true,
  linkcolor  = black
}

\title{New results from a number  operator interpretation  of the compositeness of 
bound and resonant states} %%Plural?%%
\author[]{J.~A. Oller}
\affil[]{Departamento de F\'\i sica,
Universidad de Murcia, E-30071 Murcia,
Spain}

\begin{document}
\maketitle
\begin{abstract}
 A novel theoretical approach to the problem of the compositeness ($X$)
  of a resonance or bound state is developed on the basis of the 
expectation values of the number operators of the free particles in the continuum.  
 This formalism is specially suitable for effective field theories in which the bare elementary states are 
   integrated out but that give rise to resonance and bound states
 when implemented in nonperturbative calculations.
  We demonstrate that $X=1$ for finite-range energy-independent potentials, either regular or singular.
A non-trivial example  for an energy-dependent potential is discussed where it is shown that 
$X$ is independent of any type of cutoff regulator employed.
 The generalization of these techniques to relativistic states is developed.
 We also explain how to obtain a meaningful compositeness  with respect to the open channels
  for resonances, even if it is complex in a first turn,  by making use of suitable
 phase-factor transformations. Defining elementariness as $X=0$, 
 we derive  a new universal criterion for the elementariness of a bound state. 
 Along the same lines, a necessary condition for a resonance to be qualified as
 elementary is given. 
 The application of the formalism here developed might be of considerable practical interest.

\end{abstract}

\newpage
\tableofcontents
\newpage

%%%%%%%%%%%%%%%%%%%%%%%%%%%%%%%%%%%%%%%%%%%%%%%%%%%%%%%%%%%%%%%%%%%%%%%%%
\section{Introduction and basic definitions}
\label{ref.170929.0}

The problem we consider is to discern whether a bound  state or a resonance is elementary or composite with respect to the asymptotic states in the theory.
 E.g. a Hydrogen atom is a composite state of a proton and an electron in which these two particles are typically
separated by a much larger distance than its intrinsic sizes
(this is a simple example of what is usually referred to as a ``molecular'' state).
The Hydrogen wave function  
can be fully expressed as a Fourier transform in terms of free plane waves of 
a proton and an electron.
In turn, the proton appears as a bound state in the $P$-wave of $n\pi^+$ scattering.
However, the proton, with a charge radius less than $0.9$~fm, cannot be certainly qualified as a
``molecular'' state of a $n\pi^+$ but rather the opposite. 
Another example in this direction is the baryon $\Lambda$, a resonance that decays by
weak interactions into $p\pi^-$ and $n\pi^0$, and despite this splitting
the $\Lambda$  baryon is not a composite object made up by protons and pions.
Our work here is a continuation in the historical effort to quantify the weight of the asymptotic states
in a bound or resonance state.

Since the earlier seminal papers treating the problem of compositeness/elementariness of a  state
in terms of the asymptotic particles 
\cite{Castillejo.171020.1,amado.170930.1,salam.170930.1,ref.170928.1,ref.170928.2,ref.180318.1,lurie.170930.1},
it has come clear that this set up cannot answer all the interesting questions on this respect.
 We have in mind here the clear example of Quantum Chromodynamics (QCD),
which appeared later than all these papers  \cite{qcd.180318.1}.
According to this theory the proton is indeed  a composite object of three valence quarks,
though  the quarks are not asymptotic states because of the phenomenon of color confinement in QCD \cite{qcd.180318.2}. 
Therefore, the possible impact of underlying degrees of freedom to the actual observed spectrum
leaves  open the question whether
experiment can decide what sort of elementary particles exist \cite{ref.170928.1}. Nonetheless, 
it is a definitively settled matter once it is demonstrated
  that a bound or resonance state is not elementary in terms of the asymptotic degrees of freedom.

Let us discuss first the case of a bound state within non-relativistic quantum mechanics (NRQM)
as a prototypical example of the compositeness relation.
We will discuss later its relativistic generalization within Quantum Field Theory (QFT). 
 We follow at this stage the basic set up discussed
in Refs.~\cite{ref.170928.1,ref.170928.2}, and split the full Hamiltonian $H$ in an
unperturbed free-particle part $H_0$ plus an interaction $V$,
\begin{align}
  \label{170928.1}
  H=H_0+V~.
  \end{align}
The spectrum of the full Hamiltonian consists of the continuum states
\begin{align}
  \label{170928.2}
  H|\psi_\alpha\rangle&=E_\alpha|\psi_\alpha\rangle~,
\end{align}
 and it might contain also discrete bound states $|\psi_{n}\rangle$
 \begin{align}
   \label{170928.3}
 H|\psi_{n}\rangle&=E_{n}|\psi_{n}\rangle~.
 \end{align}
 The continuum eigenstates of $H$ are normalized to Dirac delta functions and
 the discrete ones to Kronecker deltas.

 In turn the free-particle Hamiltonian $H_0$ also contains the continuum spectrum and,
 in addition, there may be discrete states (or bare elementary ones). To fix the
 notation,
 \begin{align}
   \label{170928.4}
   H_0|\varphi_\alpha\rangle&=E_{\alpha}|\varphi_\alpha\rangle~,\\
   H_0|\varphi_n\rangle&=E_{n}|\varphi_n\rangle~,\nn
 \end{align} 
 The eigenstates of $H_0$ fulfill the completeness relation
 \begin{align}
   \label{170928.5}
   I &=\int d\alpha |\varphi_\alpha\rangle
 \langle \varphi_\alpha|+\sum_n|\varphi_n\rangle \langle \varphi_n|
   \end{align}
 with $H$ and $H_0$ sharing the same spectrum \cite{ref.170928.3,ref.170928.4}. 
 Given a bound state $|\psi_B\rangle$ of $H$ with energy $E_B$
 \begin{align}
   \label{170929.3}
   H|\psi_B\rangle&=E_B|\psi_B\rangle~,
   \end{align}
we express it in terms of the eigenstates
 of $H_0$ as
 \begin{align}
   \label{170929.1b}
   |\psi_B\rangle&=
   \int d\alpha \langle \varphi_\alpha|\psi_B\rangle |\varphi_\alpha\rangle
   +\sum_n \langle \varphi_n|\psi_B\rangle |\varphi_n\rangle~.
 \end{align}
 Since  $|\psi_B\rangle$ is  normalized to unity,  
 it follows  that
 \begin{align}
   \label{170929.2}
   \langle \psi_B|\psi_B\rangle=1=\int d\alpha |\langle \varphi_\alpha|\psi_B\rangle|^2
   +\sum_n |\langle \varphi_n|\psi_B\rangle|^2=Z+X~,
 \end{align}
 where
 \begin{align}
   \label{170929.4}
   X&=\int d\alpha |\langle \varphi_\alpha|\psi_B\rangle|^2~,\\
   \label{170929.4b}
   Z&=\sum_n |\langle \varphi_n|\psi_B\rangle|^2~.
 \end{align}
 These quantities are usually called compositeness ($X$) and
 elementariness ($Z$). 
 
 Making use of the Schr\"odinger equation, 
 written in the form $(H_0+V)|\psi_B\rangle=E_B|\psi_B\rangle$, we can
 express $X$ from  Eq.~\eqref{170929.3} as \cite{ref.170928.2}
 \begin{align}
   \label{170929.5}
   X&=1-Z=\int d\alpha\frac{|\langle \varphi_\alpha|V|\psi_B\rangle|^2}{(E_\alpha-E_B)^2}~.
 \end{align}
 Notice that the integrand in the previous equation is just the modulus square of the continuum part of the
 bound-state wave function ${\psi}_B(\alpha)$. The latter is given by
 \begin{align}
      \label{170929.6}
{\psi}_B(\alpha)&=\frac{\langle \varphi_\alpha|V|\psi_B\rangle}{E_\alpha-E_B}~.
 \end{align}
 This equation is well-known in ordinary quantum mechanics for energy-independent
 local potentials \cite{Faddeev.170929.1,Gottfried.170929.2}. Within a more general
 scenario,  Eq.~\eqref{170929.5} expresses the fact that it might not be normalized to 1
 when there are 
 elementary states ($Z\neq 0$).
 %It has a simple
% diagrammatic interpretation as given in Fig.~\ref{fig.170929.1}.\footnote{This picture 
% reflects that a free-particle state in the bound state
% propagates  according to the bound-state propagator, 
% which finally gives rise again to the free system through the coupling $\langle\varphi_\alpha|V|\psi_B\rangle.$}
%\begin{figure}
%\begin{center}
%\includegraphics[width=0.4\textwidth]{wf.eps}
%\caption{Diagrammatic representation of the wave function
%  $\langle \varphi_\alpha|\psi_B\rangle$ (indicated by the curly bracket) from Eq.~\eqref{170929.6}.
%  The two parallel solid lines are the bound-state propagator, the  filled circle is
%  the coupling $\langle\varphi_\alpha|V|\psi_B\rangle$ and the dashed lines refer to the free state
%$|\varphi_\alpha\rangle$~.}
%\label{fig.170929.1}
%\end{center}
%\end{figure}

%%%%%%%%%%%%%%%%%%%%%%%%%%%%%%%%%%%%%%%%%%%%%%%%%%%%%%%%%%%%%%%%%%
\section{A different perspective on the compositeness of a bound state}
\label{ref.170929.1}

We now offer  a reinterpretation of the concept of compositeness $X$ introduced in the
previous section.
This allows one to calculate $X$  by focusing  entirely on the free particle spectrum,
which is certainly the one always accessible in scattering/production experiments,
without the need to introduce the bare-elementary state contribution to the normalization to
1 of the bound (resonance) state.

 Our main motivation here lies in the fact that in many applications within
 effective field theory (EFT),  according to our own experience,
 the bare elementary discrete states are typically integrated out and do not
 appear explicitly in the Lagrangian of the theory (which is written in terms of ``low-energy'' effective degrees of freedom).
 Nonetheless, one can still generate bound states and resonances
 after complementing the perturbative calculations in the corresponding EFT with nonperturbative techniques.
 Some examples in this respect can be found e.g. in
 Refs.~\cite{Weinberg.170929.1,kaiser.170929.1,oller.170929.1,oller.211116.5,oller.170929.2,pelaez.170929.1,colangelo.170929.1}.
 In particular, a near-threshold bare elementary discrete state can be mimicked by including a
 Castillejo-Dalitz-Dyson pole  \cite{Castillejo.171020.1} in the scattering amplitude of the free continuum states. Explicit examples are
 worked out in Refs.~\cite{oller.211116.5,kang.170930.1}.
 It is also the case that $H$ might be expressed in terms of degrees of freedom that are not asymptotically free, as it
occurs in Quantum Chromodynamics in ordinary conditions.
Therefore, trying to  calculate the wave-function renormalization factor $Z$ is not practical in such situations 
and we better
derive results from the knowledge of the scattering operator $T$ among the effective degrees of freedom.
Of course, in the situations that fit the scheme presented in Sec.~\ref{ref.170929.0}
 one could calculate $Z$ as explained there or pose the problem in the terms that we expose next.
 
%In all these cases it is not a priori clear whether an
%eigenstate of the full Hamiltonian is composite or elementary with regards to the asymptotic states in the continuum.

The new perspective  on $X$ heavily relies on the number operator for a given particle species, a basic concept in QFT \cite{ref.170928.4}.
%Let us consider the annihilation and creation  operators of the continuum eigenstates of $H_0$.
For definiteness, let us take two particle species $A$ and $B$ whose annihilation/creation
operators are denoted by  $a_\alpha/a_\alpha^\dagger$ and $b_\beta/b_\beta^\dagger$, respectively.
In terms of them $H_0$ reads
\begin{align}
   \label{170929.7}
  H_0&=\int d\alpha E_\alpha \,a_\alpha^\dagger a_\alpha
  +\int d\beta E_\beta \,b_\beta^\dagger b_\beta+\sum_n E_n|\varphi_n\rangle\langle\varphi_n|~.
\end{align}

The decomposition of the bound state in eigenstates of $H_0$, Eq.~\eqref{170929.1b}, reads now
\begin{align}
   \label{170929.8}
   |\psi_B\rangle&=
   \int d\gamma \langle AB_\gamma|\psi_B\rangle |AB_\gamma\rangle
   +\sum_n \langle \varphi_n|\psi_B\rangle |\varphi_n\rangle~.
 \end{align}

%Within our approach the number operators for the free-particle states are going to play a prominent role.
For a given particle species $A$ its number operator is denoted by $N_D^A$ and defined by
\begin{align}
  \label{170929.9}
  N_D^A&=\int d\alpha \,a_\alpha^\dagger a_\alpha~.
\end{align}
Here the subscript $D$ refers to the Dirac or interaction image. Notice that since $N_D$ and $H_0$
obviously commute then
\begin{align}
  \label{170929.10}
  N_D^A(t)=e^{i H_0 t} N_D^A(0) e^{-i H_0 t}=N_D~.
  \end{align}
Based on the number operators of $A$ and $B$  we define the compositeness $X$ of the bound state $|\psi_B\rangle$ as
\begin{align}
  \label{170929.11}
  X&=\frac{1}{2}\langle \psi_B|N_D^A+N_D^B|\psi_B\rangle~.
\end{align}
That is, $X$ is the expectation value of the number operators of the
 free-particle constituents in the eigenstate $|\psi_B\rangle$ of $H$ divided by
their nominal  number, which in this case is 2.

We can see that the new definition of $X$ is equivalent to the original one of Eq.~\eqref{170929.4}
 because 
% For that it is enough to notice the trivial result  
 $(N_D^A+N_D^B)|AB_\gamma\rangle=2 |AB_\gamma\rangle$ and
 the annihilation operators $a_\alpha$ and $b_\beta$
 destroy the bare elementary discrete states present  in Eq.~\eqref{170929.8}. 
 It then follows that  $X$, as defined in 
 Eq.~\eqref{170929.11}, reads 
\begin{align}
  \label{170929.12}
  X&=\int d\gamma |\langle AB_\gamma|\psi_B\rangle|^2~,
\end{align}
 as in Eq.~\eqref{170929.4}.

In general, if we are applying NRQM to a bound state $|\psi_B\rangle$ of 
$n$ particles corresponding to $m$ particle species $A_1$, $\ldots$, $A_m$,
the compositeness is defined by a straightforward generalization of
the two-body case of Eq.~\eqref{170929.11} as 
\begin{align}
  \label{170929.13}
  X&=\frac{1}{n}\langle \psi_B|\sum_{i=1}^m N_D^{A_i}|\psi_B\rangle~.
\end{align}
To simplify the notation in the following the sum over the number operators is
denoted simply by $N_D$
\begin{align}
  \label{170930.1}
  N_D&=  \sum_{i=1}^m N_D^{A_i}~.
  \end{align}

It is worth stressing that for a given total Hamiltonian $H$ with regular interactions $V$
  (it is sufficient that it fulfills in momentum space
 the Eqs.~\eqref{180321.1} and \eqref{180321.2} below \cite{ref.171001.2})
  the compositeness $X$ is 
  an  observable (this is similarly expressed at the end of Sec.IV of Ref.~\cite{ref.170928.1}).
 Accordingly to the postulates of NRQM this is clear  from its new interpretation in Eq.~\eqref{170929.13}
   as the expectation value of a linear self-adjoint operator. 
  The same comment also applies over partial compositeness coefficients.
  
  However, for more singular interactions a nonperturbative regularization process is required, which is an
issue that is not fully settled (we elaborate more on this point in Sec.~\ref{sec.170929.2}).
In this regards, so as to appreciate the limitation of straightforward extrapolations
 of perturbative results in renormalization theory
to nonperturbative calculations, it is written in Ref.~\cite{ref.170928.2} that generally $Z^{-1}$ is divergent in (relativistic) 
 QFT calculations, but it seems reasonable to expect that this is a failure of perturbation theory, and not
that $Z$ is really zero for all particles. We show in Sec.~\ref{sec.180321.1} that the total compositeness $X$ is one 
for a general finite-range energy-independent potential, 
irrespectively of whether it is regular or singular. 
We also give two interesting examples  in which $X$ is independent of the type of 
 cutoff regularization employed for energy-dependent potentials. 

 %%%%%%%%%%%%%%%%%%%%%%%%%%%%%%%%%%%%%%%%%%%%%%%%%%%%%%%%%%%%%%%%%%%%%%%%%%%%
\section{Quantum Field Theory calculation of $X$}
\label{sec.170929.2}

An interesting consequence of the new definition for $X$, Eq.~\eqref{170929.13},
is that it is amenable to a direct computation  within NR Quantum Field Theory (QFT).
To show it let us consider the Dirac or interaction picture and introduce the interaction adiabatically
\begin{align}
  \label{170930.19}
  V\to V e^{-\vep |t|}
\end{align}
with $\vep\to 0^+$.
 At time $t$ the states $|\varphi(t)\rangle$ in the Dirac picture are related to the
states  $|\psi\rangle$  in the Heisenberg picture by
\begin{align}
  \label{170929.14}
  |\varphi(t)\rangle &=e^{iH_0t}e^{-iHt}|\psi\rangle=U_D(t,0)|\psi\rangle~,\\
  |\varphi(0)\rangle &= |\psi\rangle~.\nn
\end{align}
The time evolution operator for the Dirac states is
denoted by $U_D(t_2,t_1)$ and it corresponds to 
$U_D(t_2,t_1)=e^{iH_0 t_2}e^{-iH(t_2-t_1)}e^{-iH_0 t_1}$.
In particular, the bound state $|\psi_B\rangle=|\varphi_B(0)\rangle$  can be expressed  by the
time evolution from the asymptotic  bare elementary discrete state $|\varphi_B\rangle$ as
\begin{align}
  \label{170929.15}
  |\psi_B\rangle&=U_D(0,-\infty)|\varphi_B\rangle~,\\
  |\psi_B\rangle&=U_D(0,+\infty)|\varphi_B\rangle~.\nn
  \end{align}
In this way
\begin{align}
  \label{170930.1b}
  X&=\frac{1}{n}\langle \varphi_B|U_D(+\infty,0)N_D U_D(0,-\infty)|\varphi_B\rangle~,
\end{align}
where we have used that the M\"oller matrix $U_D(0,+\infty)$  satisfies that 
$U_D(0,+\infty)^\dagger U_D(0,+\infty)=I$ \cite{Schweber.170930.1}.
The previous matrix element can be written in a time-ordered way by
introducing an extra time evolution from 0 to $t$. For that let us notice
that
\begin{align}
  \label{170930.2}
  |\varphi_B(t)\rangle&=U_D(t,0)|\psi_B\rangle
  =e^{iH_0 t}e^{-i Ht}|\psi_B\rangle
  =e^{iH_0 t}e^{-iE_B t} U_D(0,\pm \infty)|\varphi_B\rangle~.
\end{align}
Therefore, equating the last step with the second one, we can express
\begin{align}
\label{180603.1}
U_D(0,-\infty)|\varphi_B\rangle&=e^{i E_B t}e^{-i H_0 t} U_D(t,0)|\psi_B\rangle
=e^{i E_B t}e^{-i H_0 t} U_D(t,-\infty)|\varphi_B\rangle
\end{align}
and similarly, 
\begin{align}
\label{180603.2}
\langle \varphi_B|U(+\infty,0)&=\langle \varphi_B|U_D(+\infty,t) e^{i H_0 t} e^{-i E_B t}~.
\end{align}
Next, we replace Eqs.~\eqref{180603.1} and \eqref{180603.2} into Eq.~\eqref{170930.1b} 
which then reads
\begin{align}
\label{180603.3}
X&=\frac{1}{n}\langle \varphi_B|U_D(+\infty,t)e^{i H_0 t} e^{-i E_B t} N_D 
e^{i E_B t}e^{-i H_0 t} U_D(t,0)|\psi_B\rangle
\end{align}
for arbitrary $t$. 
The  phase factors $e^{\pm iE_B t}$ cancel out  while
\begin{align}
  \label{170930.3}
  e^{iH_0 t}N_D e^{-i H_0 t}=N_D(t)=N_D~,
    \end{align}
 recall Eq.~\eqref{170929.10}. In this way, after averaging in $t$, the 
 Eq.~\eqref{170930.1b} becomes
\begin{align}
    \label{170930.4}
    X=\frac{1}{n}\lim_{T\to +\infty}\frac{1}{T}\int_{-T/2}^{+T/2} dt \langle \varphi_B|U_D(+\infty,t)
    N_D(t)U_D(t,-\infty)|\varphi_B\rangle~,
\end{align}
which is the form that we are seeking for. The factor $1/T$ in the previous equation
cancels in the limit $T\to +\infty$ with the Dirac delta function of total energy
conservation (times $2\pi$).

It might be advantageous to express the number operator in terms of NR fields in
Eq.~\eqref{170930.4}, e.g. in order to apply Feynman diagrams for its calculation.
 For a generic scalar particle species $A_i$ of physical mass $m_{A_i}$
we have the free field
\begin{align}
  \label{171110.1}
  \psi_{A_i}(x)&=\int \frac{d^3 \vq}{(2\pi)^3} a_i(\vq) e^{-i \tilde{q} x}~,
  \end{align}
where $q^0=\vq^2/2 m_{A_i}$, $\tilde{q}=(q^0,\vq)$ and $x=(t,\vx)$~.
It is then straightforward to show that
\begin{align}
  \label{171110.2}
N_D&=\sum_i \int d^3\vx\, \psi^\dagger_{A_i}(x)\psi_{A_i}(x)~.
\end{align}
Inserting this expression into Eq.~\eqref{170930.4} it reads
\begin{align}
  \label{171110.3}
  X&=\frac{1}{n}\lim_{T\to +\infty}\frac{1}{T}\int_{-T/2}^{+T/2} dt\int d^3 \vx
 \langle \varphi_B|U_D(+\infty,t) \sum_i \psi_{A_i}^\dagger(x)\psi_{A_i}(x)
U_D(t,-\infty)|\varphi_B\rangle~,\nn\\
 &=\frac{1}{n}\lim_{T\to +\infty}\frac{1}{T}\int d^4x 
\langle \varphi_B| P\left[ e^{-i\int_{-\infty}^{+\infty}dt' V_D(t')}
 \sum_i \psi_{A_i}^\dagger(x)\psi_{A_i}(x)
 \right] |\varphi_B\rangle~.
  \end{align}
Here we denote the time-ordered product by $P$ and $V_D(t)$ is the interaction in the
Dirac picture.\footnote{In Eq.~\eqref{171110.3} only the connected diagrams should be considered
  \cite{Low.171110.1}.} 
The extension to particles with other spin is straightforward.

\begin{figure}
\begin{center}
\includegraphics[width=0.5\textwidth]{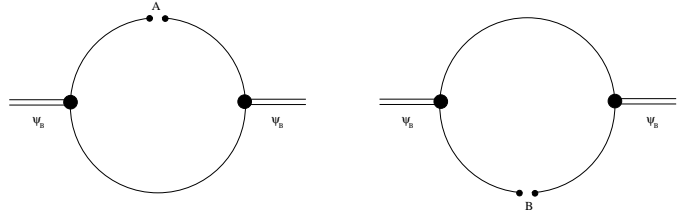}
\caption{Feynman diagrams for the calculation of $X$ within NR QFT
  for the two-particle case. The insertion of the number operators for
the particles $A$ and $B$ is indicated by the double dots.}
\label{fig.170930.1}
\end{center}
\end{figure}

For the two-particle case, with particles of types $A$ and $B$,
the evaluation of $X$ according to
Eq.~\eqref{170930.4} corresponds to  the calculation of the diagrams
in Fig.~\ref{fig.170930.1}. 
 Its evaluation is straightforward\footnote{One could make use of standard Feynman rules in 
 relativistic QFT, integrate over
 the temporal component of the loop momentum (taking into account that the coupling squared does not depend 
 on this integration variable in the NR case) and proceed with 
 the non-relativistic reduction of the kinematics. 
 The necessary steps in QFT from the Feynman diagrams in 
Fig.~\ref{fig.170930.1} are given in the Appendix \ref{app.170928.1}.
 \label{footnote.180603.1}} and
in the $\ell S$ basis (with $\ell$ the orbital angular momentum and
$S$ the total spin) we have
\begin{align}
  \label{170930.5}
  X&=\sum_{\ell,S} X_{\ell S}~,\\
  \label{170930.6}
  X_{\ell S}&=\frac{1}{2\pi^2}\int_0^\infty dk k^2 \frac{g_{\ell S}^2(k^2)}{(k^2/2\mu-E_B)^2}~.
\end{align}
In this equation, $\mu$ is the reduced mass of particles $A$ and $B$, and 
$g_{\ell S}^2(k^2)$ is the coupling  squared of the
bound state, $\langle AB_{\ell S}|V|\psi_B\rangle^2$.
This equation is in agreement with Eq.~\eqref{170929.5}.
 We discuss explicitly in Appendix \ref{app.170928.1} the
angular momentum algebra needed to express $X$ as the
diagonal sum over the compositeness in a partial wave ($X_{\ell S}$),  Eq.~\eqref{170930.5}.

 The coupling can be calculated by taking into account the Lippmann-Schwinger (LS) equation
  in partial waves
\begin{align}
  \label{170930.7}
%  T(E)=V-V\frac{1}{H_0-E}T(E)
  T(E)=V+V\frac{1}{E-H_0}T(E)
\end{align}
for the off-shell  $T$ matrix, with matrix elements $T(k',k;E)$ (if several partial
waves mix the previous LS equation is still valid in a matrix notation).
The $T$ matrix  has a pole at $E=E_B$ and then, by taking
the limit $E\to E_B$ in the LS equation,  it follows that $g(k)$ satisfies
 a homogeneous integral equation for $k\in [0,\infty]$  (again a
matrix notation should be employed if appropriate)
\begin{align}
  \label{170930.8}
%  g(k)&=\frac{1}{2\pi^2}\int_0^\infty dk'\,{k'}^2V(k,k')\frac{1}{{k'}^2/2\mu-E_B}g(k')~.
  g(k)&=\frac{1}{2\pi^2}\int_0^\infty dk'\,{k'}^2V(k,k')\frac{1}{E_B-{k'}^2/2\mu}g(k')~.
\end{align}
 From Eq.~\eqref{170930.8} and the fact that $V(-k,k')=(-1)^\ell V(k,k')$ (parity conservation)
 one concludes that the coupling squared only depends on $k^2$, as already expressed in
 Eq.~\eqref{170930.6}. 

  The global normalization factor in Eq.~\eqref{170930.8}
  is fixed by the requirement that $g(k)$
  matches the residue of the $T$ matrix at the pole position 
  \begin{align}
  \label{170930.9}
  g^2(\varkappa^2)&=\lim_{E\to E_B}(E-E_B)T(\varkappa,\varkappa;E)~.
%  g_{\ell S}g_{\ell'S'}(\varkappa^2)&=\lim_{E\to E_B}(E-E_B)T(E)_{\ell S;\ell' S'}(\varkappa,\varkappa)~,
  \end{align}
%  with the subscripts indicating the corresponding partial waves.
 Here,  $\varkappa=\sqrt{2\mu E_B}$
  with ${\rm Im}\varkappa>0$ [1st or physical Riemann sheet (RS)].

Taking advantage of the fact that the integrand in Eq.~\eqref{170930.6}
is an even function of $k$ one can symmetrize it and rewrite Eq.~\eqref{170930.6} as
\begin{align}
  \label{170930.10}
  X&=\left(\frac{\mu}{\pi}\right)^2\int_{-\infty}^{+\infty}dk k^2
  \frac{g^2(k^2)}{(k^2-\varkappa^2)^2}~,
\end{align}
where for briefing the writing we have suppressed the subscript $\ell S$.

%%%%%%%%%%%%%%%%%%%%%%%%%%%%%%%%%%%%%%%%%%%%%%%%%%%%%%%%%%%%%%%%%%%%5
\section{Calculations of $X$ in NR QFT}
\label{sec.170930.1}

We stress that there are no further contributions beyond Eq.~\eqref{170930.6} for $X_{\ell S}$. 
 The absence of tad-pole like contributions within an appropriate regularization procedure
 in NR QFT drives to Eq.~\eqref{170930.6} as the final expression without any possible counterterm
 contributions. As shown in Eq.~\eqref{170929.5} this expression also follows from
 the Schr\"odinger equation \cite{ref.170928.2}. 
 Thus, $X$ is a fully derived quantity from the knowledge of the (full or half) off-shell $T$ matrix which allows also
 to determine the coupling function $g^2(k^2)$. 
 This is an expected result because the $T$ matrix must comprise all the spectroscopical
information of the corresponding quantum system.

 The analytical properties of $g(k)$ in the
$k$-complex plane can be deduced from Eq.~\eqref{170930.8}. In the case of a separable
potential the deduction is straightforward. For more involved potentials 
one could use the techniques derived in Refs.~\cite{entem.170930.1,entem.170930.2} for the more complicated   
problem of establishing the LS equation for complex momenta  
(which are suitable for potentials that can be given by a spectral decomposition with
spectral functions analytic in $E$). This latter case 
will be explained in more detail in Ref.~\cite{entem.170930.2},
and then $g(k)$ is analytic in $k$ without cuts.
As a result, a non-constant $g(k)$ is not bounded for $k\to \infty$ in the $k$ complex plane 
because of the Liouville's theorem in complex analysis.

\begin{figure}
\begin{center}
  \includegraphics[width=0.2\textwidth]{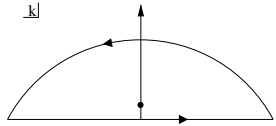}
\caption{Closed integration contour for the evaluation of the integral in Eq.~\eqref{170930.10}}
\label{fig.170930.2}
\end{center}
\end{figure}

%%%%%%%%%%%%%%%%%%%%%%%%%%%%%%%%%%%%%%%%%%%%%%%%%%%%%%%%%%%%%%%%%%%%%%%%%%%%%%%%%%%
\subsection{Zero-range potentials}
\label{sec.180319.1}

We consider first the case in which  the wavelengths of the
two scattering particles are large compared with the range of their interaction.
 In configuration space the potential is then approximated as the sum of
 delta functions  at the origin and derivatives of them 
\cite{phillips.180319.1,kolck.171107.1}.
 We treat the potential given as exact and a full nonperturbative solution of the
 LS equation is worked out in partial waves. The solution obtained
  % of the LS equation
 is also valid for energy-dependent potentials. 
 
 Despite the fact that solutions for the on-shell $T$ matrix of the LS equation differ between
 cutoff and dimensional regularization when implemented nonperturbatively \cite{phillips.180319.1}, 
 we are able to show that, regardless of the regularization method
 employed,  $X=1$  for zero-range  energy-independent potentials.

 For a given set of quantum numbers the potential for $n$ two-body coupled channels 
 (here we do not distinguish between channel and partial wave since they can be treated on the same foot) is given by the sum 
 \begin{align}
   \label{180319.1}
v_{\alpha\beta}(k_\alpha,p_\beta)&=k_\alpha^{\ell_\alpha}p_\beta^{\ell_\beta} \sum_{i,j}^N v_{\alpha\beta;ij}k_\alpha^{2i}p_\beta^{2j}~,
 \end{align}
  where the channels are numbered by  employing Greek letters. 
 The factor in front of the sum is driven by the threshold-behavior of the different partial waves, being $k_\alpha$ and $p_\beta$
 the three-momenta of the corresponding channels.
 The coefficients $v_{\alpha\beta;ij}$ could be energy dependent but they do not depend
 on the three-momenta. These coefficients  are grouped as the matrix elements of the $N\times N$ matrices 
 $[v_{\alpha\beta}]$ which, in turn, are the block matrix elements of the potential matrix
 $[v]$
 \begin{align}
   \label{180319.2}
   [v]&=\left(\begin{matrix}
     [v_{11}] & [v_{12}] & \ldots & [v_{1n}]\\
     [v_{21}] & [v_{22}] & \ldots & [v_{2n}]\\
     \ldots  & \ldots  &\ldots & \ldots \\
     [v_{n1}] & [v_{n2}] & \ldots & [v_{nn}]\\
     \end{matrix}
   \right)~.
 \end{align}
 We also introduce the $N  n$ column vectors $[k_\alpha]$ as
 \begin{align}
   \label{180319.3}
   [k_\alpha]^T&=(
   \underbrace{0,\ldots,0}_{N(\alpha-1) \text{~places}},k^{\ell_\alpha},k^{\ell_\alpha+2},\ldots,k^{\ell_\alpha+2N},0,\ldots,0)~,
 \end{align}
 such that $v_{\alpha\beta}(k_\alpha,p_\beta)$ in Eq.~\eqref{180319.1} can be conveniently written in matrix notation as
 \begin{align}
   \label{180319.4}
   v_{\alpha\beta}(k_\alpha,p_\beta)&=[k_\alpha]^T\cdot [v]\cdot [p_\beta]~.
 \end{align}
 In this way the solution for the LS equation Eq.~\eqref{170930.7}, $t_{\alpha\beta}(k,p;E)$, is also written in  matrix notation as
 \begin{align}
   \label{180319.5}
   t_{\alpha\beta}(k_\alpha,p_\beta;E)&=[k_\alpha]^T\cdot [t(E)]\cdot [p_\beta]~,
 \end{align}
 where we have introduced the scattering matrix $[t(E)]$.
 In order to satisfy the LS equation the latter is required to fulfill
 \begin{align}
   \label{180319.6}
         [t(E)]&=[v(E)]-[v(E)]\cdot [G(E)] \cdot [t(E)]~,
 \end{align}
 where the block-diagonal matrix of unitarity one-loop functions $[G(E)]$ is defined as
\begin{align}
   \label{180319.6b}
   [G(E)]&=\sum_\alpha [G_\alpha(E)]~.
   \end{align}
 The  $[G_\alpha(E)]$ is an $N n\times N n$ matrix  
\begin{align}
   \label{180319.7}
  [G_\alpha(E)]&=\frac{m_\alpha}{\pi^2}
 \int_0^\infty dq \frac{q^2}{q^2-2m_\alpha E} [q_\alpha] \cdot [q_\alpha]^T~,
   \end{align}
 with $m_\alpha$  the reduced mass of the $\alpha$ channel.  
 The algebraic solution of Eq.~\eqref{180319.6} is
 \begin{align}
   \label{180319.8}
         [t(E)]&=[D(E)]^{-1}~,\\
         [D(E)]&=[v(E)]^{-1}+[G(E)]~,
%    \label{180319.8b}
%         t_{\alpha\beta}(k_\alpha,p_\beta;E)&=[k_\alpha]^T\cdot [t(E)]\cdot [p_\beta]~.
 \end{align}
 Of course the matrices $[t(E)]$ and $[D(E)]$ have a block matrix from analogous to that of $[v]$  in Eq.~\eqref{180319.2}. 
 One  can work out several expressions for calculating $g^2_\alpha(k^2)$, which is the square of the coupling function
 of the bound state to the channel $\alpha$.
 We derive here two of them  making use of the half-off-shell $T$ matrix
 (with $E_B=-\gamma_\alpha^2/2m_\alpha$ for all the channels) that will be used below.
  In these expressions the determinant of the matrix $[D]$ cancels.

 We write  $[D]^{-1}$ in terms of the adjoint matrix of $[D]$ ($[d]=\text{adj}[D]$)  and its determinant $\Delta$,
 \begin{align}
   [D]^{-1}&=\frac{[d]}{\Delta}~.
   \end{align}
 Next, taking the limit $E\to E_B$ (we assume that the zero of $\Delta(E)$ at $E=E_B$ is of order 1)
  the residue of the half-off-shell $T$ matrix ($p_\beta=i\gamma_\beta$) provides us with the coupling functions 
 \begin{align}
   \label{180319.10}
   g_\alpha(k_\alpha)g_\beta(p_\beta)&=\lim_{E\to E_B}(E-E_B)t_{\alpha\beta}(k_\alpha,p_\beta;E)
   =\left.\frac{[k_\alpha]^T\cdot [d]\cdot [p_\beta]}{\Delta'}\right|_{E=E_B,p_\beta=i\gamma_\beta}~,\\
   \Delta'&=\left.\frac{\partial \Delta}{\partial E}\right|_{E=E_B}~.\nn
 \end{align}
 Therefore, by squaring the previous expression we have for $  g_\alpha^2(k_\alpha^2)$  
 \begin{align}
   \label{180320.1}
   g_\alpha^2(k_\alpha^2)&=\left. \frac{1}{(\Delta')^2} \frac{([k_\alpha]^T\cdot [d] \cdot [p_\beta])^2}{g^2_\beta(p_\beta^2)}
   \right|_{E=E_B,p_\beta=i\gamma_\beta}~.
 \end{align}
 Another expression is obtained by taking the derivative of $t_{\alpha\beta}(k_\alpha,p_\beta;E)$ with respect to $E$ and
 then moving to the pole position, such that  $E\to E_B$ and $p_\beta\to i\gamma_\beta$. We make use here of the result
 \begin{align}
   \frac{\partial [D]^{-1}}{\partial E}&=-[D]^{-1}\cdot \frac{\partial [D]}{\partial E} \cdot [D]^{-1}~,
 \end{align}
 that follows trivially by taking the derivative of $[D][D]^{-1}=I$. 
 We end with a double pole whose coefficient is 
 \begin{align}
   \label{180320.2}
   g_\alpha(k_\alpha)g_\beta(p_\beta)=\left. \frac{1}{(\Delta')^2}[k_\alpha]^T\cdot
   [d]\cdot \frac{\partial [D]}{\partial E}\cdot [d]\cdot [p_\beta]\right|_{E=E_B,p_\beta=i\gamma_\beta}
 \end{align}
 We combine Eqs.~\eqref{180320.1} and \eqref{180320.2} (the latter particularized on-shell, $k_\alpha\to p_\alpha= i\gamma_\alpha$),
 and express $g^2_\alpha(k_\alpha^2)$ as
 \begin{align}
   \label{180320.3}
 g^2_\alpha(k_\alpha^2)&=\left.\frac{([k_\alpha]^T\cdot [d] \cdot [p_\alpha])^2}{[p_\alpha]^T\cdot
   [d]\cdot \frac{\partial [D]}{\partial E}\cdot [d]\cdot [p_\alpha]}\right|_{E=E_B,p_{\alpha}=i\gamma_{\alpha}}
 \end{align}
 in which the factors $\Delta'$ have cancelled out. 

 We are now ready to calculate the compositeness $X_\alpha$, Eq.~\eqref{170930.6}, which now becomes
 \begin{align}
   \label{180320.4}
   X_\alpha&= \left.\frac{m_\alpha/\pi^2}{[p_\alpha]^T\cdot
  [d]\cdot \frac{\partial [D]}{\partial E}\cdot [d]\cdot [p_\alpha]}\frac{\partial}{\partial E}\int_0^\infty dk \frac{k^2}{k^2-2m_\alpha E}
[k_\alpha]^T\cdot [d] \cdot [p_\alpha] \, [k_\alpha]^T\cdot [d] \cdot [p_\alpha]
\right|_{E=E_B,p_\alpha=i\gamma_\alpha}
   \end{align}
 
 Let us assume that $[v]$ is energy independent, $\partial [v]/\partial E=0$. Then, we have that
 \begin{align}
   \label{180320.5}
   \frac{\partial [D]}{\partial E}&=
\frac{\partial [G]}{\partial E}~,   
 \end{align}
 which is a block-diagonal matrix with
 \begin{align}
   \label{180320.6}
\frac{\partial [G_\alpha]}{\partial E}&=\frac{m_\alpha}{\pi^2}\frac{\partial}{\partial E}\int_0^\infty dk \frac{k^2}{k^2-2m_\alpha E} [k_\alpha]\cdot [k_\alpha]^T
 \end{align}
 For $\partial [v]/\partial E=0$ the total compositeness $X=\sum_\alpha X_\alpha=1$,
 as can be seen by performing the following steps in Eq.~\eqref{180320.4}:

 First, we rewrite the denominator in the first fraction
 of this equation by taking into account Eq.~\eqref{180320.6} as
 \begin{align}
   \label{180320.7}
   \sum_\beta [p_\alpha]^T\cdot [d] \cdot \frac{m_\beta}{\pi^2}\frac{\partial}{\partial E}\int_0^\infty dk \frac{k^2}{k^2-2m_\beta E} [k_\beta]\cdot [k_\beta]^T\cdot [d] \cdot [p_\alpha]  
 \end{align}

 Second,  we make use of Eq.~\eqref{180319.10} and rewrite the factors $[p_\alpha]^T\cdot [d] \cdot [k_\beta]$ as
 $\Delta '(E_B) g_\alpha(p_\alpha) g_\beta(k_\beta)$, and similarly for $[k_\beta]^T\cdot  [d] \cdot [p_\alpha]$.\footnote{Because
   of time reversal symmetry $[d]$ is a symmetric matrix.} It follows that,
 \begin{align}
   \label{180320.8}
   \sum_\alpha 
 X_\alpha&=\sum_{\alpha=1}^n \left.\frac{g_\alpha^2(p_\alpha^2) \frac{m_\alpha}{\pi^2}\frac{\partial}{\partial E}\int_0^\infty dk\frac{k^2}{k^2-2m_\alpha E}g_\alpha^2(k_\alpha^2)}{g_\alpha^2(p_\alpha^2)\sum_\beta
   \frac{m_\beta}{\pi^2}\frac{\partial}{\partial E}\int_0^\infty dk\frac{k^2}{k^2-2m_\beta E}g_\beta^2(k_\beta^2)}
 \right|_{E=E_B,p_\alpha=i\gamma_\alpha}\nn\\
 &=1~.
\end{align}

 Now that we have shown that the total compositeness is one for zero-range energy-independent potentials it is illustrative to
 explicitly calculate $X$ employing a particularly simple regularization method.
 In this way we also show the emergence of other contributions to $X$,
 beyond the prototypical Weinberg's result of Ref.~\cite{ref.170928.2} in the limit of vanishing
 binding energy.
 A simple way to treat with the power-like divergences that emerge when employing a potential like that
 in Eq.~\eqref{180319.1} is to regularize the potential as
\begin{align}
  \label{171007.1}
  V(k',k)\to V(k',k) e^{i\epsilon (k+k')}~.
\end{align}
with $\epsilon \to 0^+$. 
The use of the convergent factor $e^{i\epsilon (k+k')}$ removes  all the power-like divergences at the same time that it
preserves the right analytical properties. Indeed, it gives the same results
as dimensional regularization for three dimensions (as it is our case here).
 The redefinition of $V(k',k)$ in Eq.~\eqref{171007.1}
 transforms $g(k)$ as 
\begin{align}
  \label{171007.2}
g(k) \to g(k) e^{i\epsilon k}~,
\end{align}
as it is clear from Eq.~\eqref{170930.8}. 
The presence of $e^{i\epsilon k}$ in the coupling function 
allows us to  close the integration contour of the integral in  Eq.~\eqref{170930.10} along the
upper half plane of the $k$-complex plane with a semicircle of infinite radius,
as shown in Fig.~\ref{fig.170930.2}.
%The upper half plane of the complex $k$ plane
%is the one chosen for commodity because
%when $k$ is expressed in terms of the energy the bound-state pole lies in the 1st RS, with
%${\rm Im}k>0$.\footnote{Nonetheless, since $g^2(k^2)$ is a function of $k^2$ one could also
%  close the integration contour along the lower half plane.}
%Therefore, the most natural procedure is to think of $g(\varkappa)$, even though its
%square only depends on $k^2$, which also allows one to close the integration contour along the  and then must be employed because
%the bound state is a pole at the 1st RS. 
The calculation is straightforward by applying the Cauchy's integration theorem
with the result
\begin{align}
  X&=\frac{2i\mu^2}{\pi}\frac{\partial}{\partial k}
  \left[\frac{k^2 g^2(k^2)}{(k+\varkappa)^2}\right]_{k=\varkappa}\nn\\
  &=g^2(\varkappa^2)\frac{\mu^2}{2\pi \gamma}
  +\frac{\mu^2}{2\pi}\left.\frac{\partial g^2(-\bar\gamma^2)}{\partial \bar\gamma}\right|_{\bar\gamma={\gamma}}~.
  \label{170930.11}
\end{align}
where $\bar\gamma=-ik$ and $\gamma=-i\varkappa$. 
    The first term on the right hand side (r.h.s.) of this equation is a well-known
    contribution
%    , although obtained with less generality
 \cite{ref.170928.2,hyodo.170930.1,oset.170930.1,nagahiro.180603.1,sekihara.171002.1}.
    It is model independent in the sense that
    it is fixed once the pole position and the residue of the on-shell $T$ matrix at the pole position
  is known (which in principle can be  be fixed from experiment). 
  The second term of Eq.~\eqref{170930.11} is an extra contribution, which
  cannot be fixed directly from the knowledge of the on-shell $T$ matrix
    %experimental data
    and  depends on the interaction $V(k',k)$. For its evaluation
  one needs first to solve the integral equation for $g(k)$, Eq.~\eqref{170930.8}. The next step is
  to proceed by analytical continuation and evaluate the derivative of $g(k)$
  at $k=\varkappa$. This extra term is sensitive to the threshold dependence $g_{\ell S}^2(k^2)\propto k^{2\ell}$.

 The first explicit example is a pure $S$-wave
 potential given by
 \begin{align}
   \label{171107.2}
V(k',k)&=\left[v_0+v_2 (k^2+{k'}^2)\right]e^{i\epsilon(k+k')}~, 
 \end{align}
 where $v_0$ and $v_2$ are constants. 
 For our purposes of calculating $g(k)$ it is enough to work out the half-off-shell $T$ matrix, 
 with $E=k^2/2\mu$. 
  The latter can be solved in the form, cf. Eq.~\eqref{180319.5}, 
 \begin{align}
\label{171107.3}
T(E)(k',k)&=\left[t_0(E)+t_2(E)(k^2+{k'}^2)\right]e^{i\epsilon(k+k')}~.
 \end{align}
 When substituting Eqs.~\eqref{171107.2} and \eqref{171107.3} in the LS equation, Eq.~\eqref{170930.7},
 one encounters the integrals
 \begin{align}
\label{171107.4}
   \lim_{\vep,\epsilon\to 0^+}
 \frac{\mu}{2\pi^2}\int_{-\infty}^\infty dq \frac{q^{2+n}e^{2i\epsilon q}}{k^2+i\vep-q^2}=-\frac{i\mu k^{n+1}}{2\pi}~,
 \end{align}
 which are calculated following the same procedure as explained 
 with regards Eq.~\eqref{170930.11}. 
 We then find
 \begin{align}
\label{171107.5}
T(k',k;E)&=\frac{v_0+v_2(k^2+{k'}^2)}{D(E)}~,\\
D(E)&=1+i\frac{\mu\sqrt{2\mu E}}{2\pi}[v_0+ 4\mu E v_2]~.\nn
 \end{align}
 From the residue of the $T$ matrix at the pole position one can calculate straightforwardly the coupling
 function $g(k)$. Its square is
 \begin{align}
   \label{171107.6}
   g^2(k^2)&=\frac{2\pi{\gamma}/\mu^2}{1-6{\gamma}^2 v_2/v_0}
   \frac{\left(1+(k^2-{\gamma}^2)v_2/v_0\right)^2}{1-2{\gamma}^2 v_2/v_0}~.
 \end{align}
 For energy-dependent $v_0$ or $v_2$  the expression for the $T$-matrix in Eq.~\eqref{171107.5}
 is still valid, since the energy $E$ enters only parametrically in the LS equation. However, the formula for
 $g^2(k^2)$ would be different.
 
 By replacing Eq.~\eqref{171107.6} for $g^2(k^2)$ into Eq.~\eqref{170930.11} we find that both terms in the
 right-hand side of the equation give rise to
 non-zero contributions which sum is 1, as they should.
 More specifically the partial contributions of the first and second terms  are 
 \begin{align}
   \label{171107.7}
   g^2(\varkappa^2)\frac{\mu^2}{2\pi{\gamma}}&=\frac{1-2{\gamma}^2 v_2/v_0}{1-6{\gamma}^2 v_2/v_0}~,\\
   \frac{\mu^2}{2\pi}\left.\frac{\partial g^2(-\bar\gamma^2)}{\partial \bar\gamma}\right|_{\bar\gamma={\gamma}}&=
 -\frac{4{\gamma}^2 v_2/v_0}{1-6{\gamma}^2 v_2/v_0}~.\nn
\end{align}
 In this case, since $g^2(k^2)$ is not zero for $k=0$ the last contribution in Eq.~\eqref{171107.7} is
 suppressed by a factor ${\gamma}^2 |v_2/v_0|\sim \bar{\gamma}R$.
 The last step is based on the relation between $v_0$ and $v_2$ with the effective range parameters.
 Being $a$ the scattering length and $r$ the effective range we have
 \begin{align}
   \label{171107.8}
   v_0&=\frac{2\pi a}{\mu}~,\\
 v_2&=\frac{\pi a^2 r}{2\mu}~,\nn
 \end{align}
 as can be easily worked out. Therefore,  $v_2/v_0=r a/4$ and then
 ${\gamma}^2 |v_2/v_0|\sim {\gamma} R/4$.
 Here, we take into account that for standard situations $r={\cal O}(R)$ \cite{Bethe.171107.1},
 a exception would be a zero of the partial wave close enough to threshold, 
 and then  for a shallow bound state it follows that ${\gamma}= 1/a+{\cal O}({\gamma}R)$.

 Next, we work out another example in which the second term on the right-hand-side
 of Eq.~\eqref{170930.11} is not suppressed compared to the first one for shallow bound states.
 This occurs when $g^2(k^2)$ is zero at $k^2=0$, in which case the derivative of $g^2(-\bar\gamma^2)$
 with respect to $\bar\gamma$ gives rise to a term that counts as $g^2(-{\gamma}^2)/{\gamma}$.
 As a specific example let us take a  potential projected with orbital angular momentum $\ell$ which
 reads
 \begin{align}
   \label{171107.9}
V(k',k)&= v_\ell {k'}^\ell k^\ell e^{i\epsilon (k+k')}~,
 \end{align}
 where $v_\ell$ is a constant (this potential is separable).
  The solution of the LS equation $T(k',k;E)$ and the coupling function is
  \begin{align}
    \label{171107.11}
    T(k',k;E)&=\frac{v_\ell {k'}^\ell k^\ell}{D(E)}~,\\
    D(E)&={1+i\frac{\mu(\sqrt{2\mu E})^{2\ell+1}}{2\pi}v_\ell}~,\nn\\
g^2(-\bar\gamma^2)&=\frac{\bar\gamma^{2\ell}2\pi}{\mu^2(2\ell+1){\gamma}^{2\ell-1}}~.\nn   
  \end{align}
  The two terms that sum up $X=1$ in Eq.~\eqref{170930.11} are, in order,
\begin{align}
  \label{171107.12}
  1&=\frac{1}{2\ell+1}+\frac{2\ell}{2\ell+1}~.
\end{align}
We see that both contributions count on the same footing (though as $\ell$ increases the 2nd one
 becomes indeed  dominant).

%%%%%%%%%%%%%%%%%%%%%%%%%%%%%%%%%%%%%%%%%%%%%%%%%%%%%%%%%%%%%%%%%%%%%%%%%%%%%%%%%%%
\subsection{Regular and singular potentials}
\label{sec.180321.1}

Analogously as in the previous section we consider a partial-wave projected potential for the coupling
of $n$ two-body coupled channels with a given set of quantum numbers. The difference is that now 
we do not assume a zero-range interaction, as in Eq.~\eqref{180319.1}, but a general finite-range potential
in coupled channels $v_{\alpha\beta}(k_\alpha,p_\beta)$. It is not necessary that this potential be local.
When the following requirements are satisfied
\begin{align}
  \label{180321.1}
  \int_0^\infty\int_0^\infty dk \,dp|v_{\alpha\beta}(k,p)|^2<\infty 
  \end{align}
and
\begin{align}
  \label{180321.2}
  \int_0^\infty dp|v_{\alpha\beta}(k,p)|^2<M~, 
\end{align}
with $M$ a bound independent of $k$, $\alpha$ and $\beta$,
the potential can be approximated with arbitrary precision as a separable potential of rank $N$, with
$N$ arbitrarily large \cite{ref.171001.2,courant.180319.1}.\footnote{This is the basis for the Schmidt method to solve 
Fredholm integral equations \cite{tricomi.180322.1}, 
extensively used to study the LS equation and its solutions in Ref.~\cite{ref.180318.1}.} 
The potential is qualified as regular then. 
If this is not the case one should introduce a regularization method (e.g. some sort of cutoff regularization)
 such that Eqs.~\eqref{180321.1} and \eqref{180321.2} are fulfilled with the regularized potential, denoted in the following as
$\omega_{\alpha\beta}(k_\alpha,p_\beta)$, and the potential is qualified as singular. 
Note that if the potential has a finite range the integrals in Eqs.~\eqref{180321.1} and \eqref{180321.2}
are finite in the lower limit of integration.

 We consider a complete set of orthonormal linearly independent real  functions
 $\{f_s(k)\}$ in $[0,\infty)$ (we could relax the condition of being real functions and allow also complex ones, but 
then the writing would be more cumbersome).
   Had we regularized $v_{\alpha\beta}$ with a sharp cutoff $\Lambda$,  such that
 $\omega_{\alpha\beta}(k_\alpha,p_\beta)=\theta(\Lambda-p_\alpha)\theta(\Lambda-k_\beta)
   v_{\alpha\beta}(k_\alpha,p_\beta)$, it would be enough that these functions be complete in
   $[0,\Lambda)$.\footnote{In this case we could use e.g. the Legendre polynomials $\{P_\ell(x)\}$ with
       $x=k/\Lambda$.} 
The potential $\omega_{\alpha\beta}(k_\alpha,p_\beta)$ is expanded in this basis of functions
     \begin{align}
       \label{180321.3}
\omega_{\alpha\beta}(k_\alpha,p_\beta)&=\sum_{s,s'=1}^\infty f_s(k_\alpha)\omega_{\alpha\beta;ss'}f_{s'}(p_\beta)~,
     \end{align}
     with the coefficients given by
     \begin{align}
\label{180321.4}
\omega_{\alpha\beta;ss'}&=\int_0^\infty \int_0^\infty dk dp f_{s}(k)\omega_{\alpha\beta}(k,p)f_{s'}(p)
\end{align}
Then, we approximate $\omega_{\alpha\beta}(k_\alpha,p_\beta)$ by a separable potential of rank $N$ \cite{ref.171001.2}, 
$\omega^{(N)}_{\alpha\beta}(k_\alpha,p_\beta)$, given by the truncation of the previous series 
in Eq.~\eqref{180321.3} 
     \begin{align}
\label{180321.5}
\omega^{(N)}_{\alpha\beta}(k_\alpha,p_\beta)&=\sum_{s,s'=1}^N f_s(k_\alpha)\omega_{\alpha\beta;ss'}f_{s'}(p_\beta)~.
\end{align}

The solutions of the LS equation for the truncated potential $\omega^{(N)}_{\alpha\beta}(k_\alpha,p_\beta)$
 is denoted by $t^{(N)}_{\alpha\beta}(k_\alpha,p_\beta;E)$, which fulfills
\begin{align}
\label{180321.6}
t^{(N)}_{\alpha\beta}(k_\alpha,p_\beta;E)&=\omega^{(N)}_{\alpha\beta}(k_\alpha,p_\beta) 
+\sum_\gamma \frac{m_\gamma}{\pi^2}\int_0^\infty dq \frac{q^2}{q^2-2m_\gamma E}\omega^{(N)}_{\alpha\gamma}(k_\alpha,q)
t^{(N)}_{\gamma\beta}(q,p_\beta;E)~.
\end{align}
Let us show that $t^{(N)}_{\alpha\beta}(k_\alpha,p_\beta;E)$ is given by the truncated series expansion of 
 $t_{\alpha\beta}(k_\alpha,p_\beta;E)$, namely,
\begin{align}
\label{180321.7}
t^{(N)}_{\alpha\beta}(k_\alpha,p_\beta;E)&=\sum_{s,s'=1}^N f_s(k_\alpha)t_{\alpha\beta;ss'}(E) f_{s'}(p_\beta)~,\\
t_{\alpha\beta;ss'}(E)&=\int_0^\infty\int_0^\infty dk dp f_s(k)t_{\alpha\beta}(k,p;E) f_{s'}(p)~.\nn
\end{align} 
Implementing the expansions of $t^{(N)}$ and $\omega^{(N)}$ in the LS equation of Eq.~\eqref{180321.6}, and 
recalling the orthonormal character of functions $\{f_s(k)\}$,  we have the 
following algebraic equation for the coefficients $t_{\alpha\beta;ss'}(E)$,
\begin{align}
\label{180321.8}
t_{\alpha\beta;ss'}(E)&=\omega_{\alpha\beta;ss'}
+\sum_{\gamma=1}^n\sum_{s'',s'''=1}^N\omega_{\alpha\gamma;ss''}
\frac{m_\gamma}{\pi^2}
\int_0^\infty dq\frac{q^2}{q^2-2m_\gamma E}f_{s''}(q)f_{s'''}(q)\,t_{\gamma\beta;s'''s'}(E)~.
\end{align}
 We now settle an analogous matrix notation to that of Sec.~\ref{sec.180319.1}. 
 The coefficients $\omega_{\alpha\beta;ss'}$ are collected in the $N\times N$ matrices $[\omega_{\alpha\beta}]$ which are the block matrix 
 elements of the matrix $[\omega]$,
 \begin{align}
   \label{180321.9}
   [\omega]&=\left(\begin{matrix}
     [\omega_{11}] & [\omega_{12}] & \ldots & [\omega_{1n}]\\
     [\omega_{21}] & [\omega_{22}] & \ldots & [\omega_{2n}]\\
     \ldots  & \ldots  &\ldots & \ldots \\
     [\omega_{n1}] & [\omega_{n2}] & \ldots & [\omega_{nn}]\\
     \end{matrix}
   \right)~.
 \end{align}
Similarly to Eq.~\eqref{180319.3} we also introduce the $N n$ column vector $[f(k_\alpha)]$ as
\begin{align}
\label{180321.10}
[f(k_\alpha)]^T&=(
   \underbrace{0,\ldots,0}_{N(\alpha-1) \text{~places}},f_1(k_\alpha),f_2(k_\alpha),\ldots,f_N(k_\alpha),0,\ldots,0)
\end{align} 
 The unitarity loop functions are gathered in the 
 block diagonal matrix $[G(E)]$, cf. Eq.~\eqref{180319.6b}, but now $[G_\alpha(E)]$ is given by 
  \begin{align}
\label{180321.11}
[G_\alpha(E)]&=\frac{m_\alpha}{\pi^2}\int_0^\infty dq\frac{q^2}{q^2-2m_\alpha E}[f_\alpha(q)]\cdot [f_\alpha(q)]^T~.
 \end{align}
Using this nation we then have
\begin{align}
\label{180322.1}
\omega_{\alpha\beta}(k_\alpha,p_\beta)&=[f(k_\alpha)]^T\cdot [\omega]\cdot [f(p_\beta)]~,\\
t_{\alpha\beta}(k_\alpha,p_\beta;E)&=[f(k_\alpha)]^T\cdot [t(E)]\cdot [f(p_\beta)]~. \nn
\end{align}
The LS equation reduces to the algebraic Eq.~\eqref{180319.6}, whose solution is the same as the one 
in Eq.~\eqref{180319.8} but now 
%$t_{\alpha\beta}(k_\alpha,p_\beta;E)$ is given by  Eq.~\eqref{180322.1}, 
% which is the same as Eq.~\eqref{180319.5}
 with $[k_\alpha]$ and $[p_\beta]$ replaced by $[f(k_\alpha)]$ and $[f(p_\beta)]$, in order. 
Indeed we can also perform the same replacements in Eq.~\eqref{180320.4} to obtain the expression for $X_\alpha$,
\begin{align}
   \label{180322.2}
   X_\alpha= \frac{m_\alpha/\pi^2}{[f(p_\alpha)]^T\cdot
  [d]\cdot \frac{\partial [D]}{\partial E}\cdot [d]\cdot [f(p_\alpha)]}&
 \frac{\partial}{\partial E}\int_0^\infty dk \frac{k^2}{k^2-2m_\alpha E}
[f(k_\alpha)]^T\cdot [d] \cdot [f(p_\alpha)] \\
 \times& [f(k_\alpha)]^T\cdot [d] \cdot [f(p_\alpha)]
\bigg|_{E=E_B,p_\alpha=i\gamma_\alpha}
\end{align}

For an energy-independent potential, 
%$\partial [v]/\partial E=0$, 
 $\partial v_{\alpha\beta}(k,p)/\partial E=0$, 
we can follow analogous steps as those in
 Eqs.~\eqref{180320.5}--\eqref{180320.8} to end with 
\begin{align}
\label{180324.17b}
\sum_{\alpha=1}^n X_\alpha&=1~.
\end{align}

We conclude that $X=1$ for regular or singular energy-independent potentials since this result
 is always the same independently of how large $N$ and $\Lambda$ are [the latter needed for
  a singular finite-range potential to satisfy Eqs.~\eqref{180321.1}
 and \eqref{180321.2}]. 
 This demonstrates that in these cases the right normalization of the bound-state wave function is to one. 
 Physically our derivation means that the total number of asymptotic particles in the continuum  of any sort involved is two. 
 Of course, this demonstration could also be used in the case of zero-range potentials, but we have preferred to be more specific for them 
  because of its intrinsic importance at the practical and conceptual level.

%%%%%%%%%%%%%%%%%%%%%%%%%%%%
\subsection{Exchange of a bare elementary particle}
\label{sec.180323.1}

Let us assume that the free Hamiltonian $H_0$ has an elementary particle eigenstate $|0\ra$,
\begin{align}
\label{180323.1}
H_0|0\ra&=E_0|0\ra~,\\
\la0|0\ra&=1~.\nn
\end{align}
As in Ref.~\cite{ref.170928.1}  we express the full $T(E)$ matrix in terms of a ``proper" $T$-operator 
$T_1(E)$ defined as what $T(E)$ would be if the elementary particle were omitted in sums over intermediate 
states. The relation between them is rather simple and intuitive \cite{ref.170928.1}
\begin{align}
\label{180323.2}
T(E)=&T_1(E)+T_1(E)|0\ra \Delta(E)\la 0|T_1(E)~,\\
\Delta(E)&=\left[E-E_0-\Pi(E)\right]^{-1}~,\nn\\
\Pi(E)&=\la 0|T_1(E)|0\ra~.\nn
\end{align}
Here we see that the total $T$-matrix is $T_1(E)$ plus an extra term coming from the exchange 
 of the elementary particle with a fully dressed propagator $\Delta(E)$, being $\Pi(E)$ 
the corresponding self-energy. Notice also that $T_1(E)|0\ra$ is the complete vertex that converts 
the virtual elementary particle into the outgoing particles.

At the pole position of the assumed bound state $E_B$ the full propagator $\Delta(E)$ vanishes, 
which implies the equation 
\begin{align}
\label{180323.3}
E_B-E_0-\Pi(E_B)&=0~,
\end{align}
that gives the relation between the unrenormalized mass $E_0$ and the physical one $E_B$. 
 The residue of the $T$-matrix between particle states in the continuum gives us the coupling functions
\begin{align}
\label{180323.4}
g_\alpha(k_\alpha)g_\beta(p_\beta)&=Z\,
\langle k_\alpha,\alpha|T_1(E_B)|0\ra \la 0|T_1(E_B)|p_\beta,\beta\ra~.
\end{align}
Since $E_B<0$ the last factor in the previous equation is the same as $\la p_\beta,\beta|T_1(E_B)|0\ra$. 
In Eq.~\eqref{180323.4} we denote by $Z$ the wave function renormalization of the 
bare elementary field, which is the residue of $\Delta(E)$ at the pole position
\begin{align}
\label{180323.5}
Z&=\left. \left[1-\frac{\partial \Pi(E)}{\partial E}\right]^{-1}\right|_{E=E_B}~.
\end{align}
For a two-body system with the quantum numbers of the elementary state, we introduce the ``bare" 
coupling constant by 
$\widetilde{g}_\alpha(k_\alpha)=\langle k_\alpha,\alpha|T_1(E_B)|0\ra=Z^{-1/2}g_\alpha(k_\alpha)$. Then the self-energy $\Pi(E)$ 
is given by 
\begin{align}
\label{180323.6}
\Pi(E)&=-\sum_\beta\frac{m_\beta}{\pi^2}\int_0^\infty 
dk\frac{ k^2 }{k^2+\gamma_\beta^2}\widetilde{g}_\beta^2(k_\beta^2)~,
\end{align}
and its derivative by
\begin{align}
\label{180323.7}
\frac{\partial \Pi(E)}{\partial E}&=-\sum_\beta \frac{2 m_\beta^2}{\pi^2}
\int_0^\infty 
dk\frac{ k^2 }{(k^2+\gamma_\beta^2)^2} \widetilde{g}_\beta^2(k^2)~.
\end{align}
Given Eqs.~\eqref{170930.6}  and \eqref{180323.4} we have for $X_\alpha$,
\begin{align}
\label{180323.8}
X_\alpha&=\frac{1}{1+\sum_\beta \frac{2 m_\beta^2}{\pi^2}
\int_0^\infty dk\frac{ k^2 }{(k^2+\gamma_\beta^2)^2}\widetilde{g}_\beta^2(k_\beta^2)}
\frac{2m_\alpha^2}{\pi^2}\int_0^\infty 
dk \frac{k^2 }{(k^2+\gamma_\alpha^2)^2} \widetilde{g}_\alpha^2(k^2)~.
\end{align}
From the previous expression it follows the basic relation, cf. Eq.~\eqref{170929.5}, 
\begin{align}
\label{180323.9}
X&=\sum_\alpha X_\alpha=1-Z~.
\end{align}

The simplest example for this scenario is that with a constant bare coupling, 
\begin{align}
\label{180323.9b}
\la 0|V|k_\alpha,\alpha\ra=\la k_\alpha,\alpha|V|0\ra=\widetilde{g}_\alpha~,
\end{align}
with all the other matrix elements involving particles in the continuum being zero. 
  For this example Eq.~\eqref{180323.8} becomes
\begin{align}
\label{180323.10}
X_\alpha&=
\frac{\widetilde{g}_\alpha^2 m_\alpha^2/(2\pi\gamma_\alpha)}{1+\sum_\beta \widetilde{g}_\beta^2 m_\beta^2/(2\pi\gamma_\beta)}~.
\end{align}
This value is independent of regulator. This can be seen by performing the renormalization of the 
on-shell $T$-matrix, from which the value of the bare coupling constant can be obtained. E.g. for the one-channel 
 case (to simply matters) we have, Eq.~\eqref{180319.8},
\begin{align}
\label{180323.11}
T(k,p;E)&=\left[\frac{1}{\widetilde{g}^2}(E-E_0)
+\frac{m}{\pi^2}\int_0^\infty dk\frac{k^2}{k^2-2m E-i\vep}\right]^{-1}~.
\end{align}
Thus, $E_0/\widetilde{g}^2$ absorbs the divergence of the  unitarity integral (which is finite after 
a subtraction is done) by renormalizing $E_0$, while  $\widetilde{g}^2$ can be determined by the energy dependence of the 
phase shifts. Notice that if we match with the effective range expansion then the effective range resulting 
from Eq.~\eqref{180323.11} should be negative because $\widetilde{g}^2\geq 0$, cf. Eqs.~\eqref{180324.5} and \eqref{180324.6} below.  

An example which explicitly gives rise to diverging integrals for $X$ and $Z$ is the same as before but with 
the bare coupling function squared proportional to $k^2$, 
\begin{align}
\label{180323.12}
\widetilde{g}(k)^2&=\lambda k^2~.
\end{align}
In this way, had we used straightforwardly this bare coupling function in the calculation of $X$ then the  integration 
\begin{align}
\label{180324.1}
\frac{2m}{\pi^2}\int_0^\infty 
dk \frac{\lambda k^4}{(k^2+\gamma_\alpha^2)^2} ~,
\end{align}
 would be divergent. However, the correct calculation of $X$ requires the complete 
 coupling function squared, for which determination we need to implement nonperturbative regularization 
 and renormalization. We show below that 
once this is accomplished the compositeness $X$ has a value independent of the type of cutoff 
regularization employed in the limit $\Lambda\to \infty$.

To calculate the $T$-matrix we apply Eq.~\eqref{180319.2} with 
\begin{align}
\label{180324.2}
[v]&=[v_{11}]=\frac{1}{E-E_0}\left(
\begin{matrix}
0&v_{12}\\
v_{12} & 0
\end{matrix}
\right)~.
\end{align}
The unitarity loops in $[G_1(E)]$ are
\begin{align}
\label{180324.3}
I_{n+1}&=\frac{m}{\pi^2}\int_0^\infty dq \frac{q^2 q^{n}}{q^2-2mE-i\vep}
\end{align}
with $n=1$, 3 or 5 in the present case. These integrals are divergent so that regularization and 
renormalization are necessary. The divergences can be identified to arise from the simpler integrals 
\begin{align}
\label{180324.4}
L_{n+1}&=\int_0^\infty dq q^n=\theta_n \Lambda^{n+1}~,
\end{align}
with $\Lambda$ some sort of cutoff, whose precise type fixes the value of the numbers $\theta_n$. 
E.g. $\theta_n=1/(n+1)$ for a sharp cutoff regularization. In the case of dimensional regularization all of them 
vanish, $\theta_n=0$ and $L_{n+1}=0$. 
%Without entering in all the details to avoid writing long expressions that departs from the main topic of the  manuscript, 
 Employing this notation,  the matrix $[G_1(E)]$, with $k=\sqrt{2 m E}$, reads
\begin{align}
[G_1(E)]&=
\frac{m}{\pi^2}\left(
\begin{array}{ll}
L_1 +i\frac{\pi}{2} k& L_3+k^2 L_1 +i\frac{\pi}{2} k^3 \\
L_3+k^2 L_1 +i\frac{\pi}{2} k^3 & L_5+k^2 L_3+L_1 k^4 +i\frac{\pi}{2} k^5 
\end{array}
\right)~.
\end{align}

We match the on-shell $T$-matrix with the effective range expansion in powers of 
$k^2$ around $k=0$, which reads
\begin{align}
\label{180324.5}
\frac{1}{T(k,k)}&=\left[\alpha+\frac{1}{2}rk^2+{\cal O}(k^4)+i\frac{mk}{2\pi}\right]^{-1}~.
\end{align}
The relation with the standard scattering length $a_s$ and effective range $r_s$ is
\begin{align}
\label{180324.6}
a_s&=\frac{m}{2\pi\alpha}~,\\
r_s&=-\frac{2\pi}{m}r~.\nn
\end{align}
For cutoff regularization in the limit $\Lambda\to \infty$ we obtain 
\begin{align}
\label{180324.7}
\frac{1}{T(k,k)}&=\alpha+\frac{1}{2}rk^2+i\frac{mk}{2\pi} ~.
\end{align}
Notice that here there is not expansion in $k^2$, so that 
 the previous result is the limit $\Lambda\to \infty$ for the 
on-shell $T$-matrix once the bare parameters $E_0$  
and $v_{12}$ are expressed as a function of $\alpha$, $r$ and $\Lambda$: 
\begin{align}
\label{180324.17}
E_0&=v_{12}\left(L_3+\epsilon\sqrt{L_5(L_1-\alpha)}\right)~,\\
v_{12}&=\frac{2\epsilon\sqrt{L_5(L_1-\alpha)}}{m\left(r L_5-2L_3(L_1-\alpha)-4\epsilon\sqrt{L_5}(L_1-\alpha)^{3/2}\right)}~,\nn
\end{align}
with $\epsilon=\pm 1$. 
We note that the potential of Eq.~\eqref{180324.2} can give rise to $r_s$ of either sign 
while keeping real values for the bare parameters $E_0$ and $v_{12}$. 
 This is not possible for the energy-independent 
 potential $v=v_{11}+v_{12}(k^2+p^2)$ because the bare parameter $v_{12}$ becomes complex for $r_s>0$, as shown in 
Ref.~\cite{phillips.180319.1}. Nonetheless, we derive below that the cutoff regularized result in the 
limit $\Lambda\to\infty$ is inconsistent for $r_s>0$  because the requirement $0\leq X\leq 1$ does not hold. 

 For the half-off-shell $T$ matrix, $T(k,p)$ we have in the same limit
\begin{align}
\label{180324.8}
\frac{T(k,p)}{T(k,k)}&=1+(k^2-p^2)\frac{\rho_\Lambda}{\Lambda^2}+{\cal{O}}(\Lambda^{-3})~,
\end{align} 
where $\rho_\Lambda$ depends on the type of cutoff regularization method employed.

In the case of dimensional regularization we obtain for the on-shell $T$-matrix
\begin{align}
\label{180324.9}
\frac{1}{T(k,k)_{DR}}&=\frac{k^2-2mE_0}{4m v_{12} k^2}+i\frac{mk}{2\pi}~,
\end{align}
that can only be matched with the effective range expansion if $E_0=0$, in which case we are left with only 
the scattering-length approximation
\begin{align}
\label{180324.10}
\frac{1}{T(k,k)_{DR}}&=\alpha+i\frac{mk}{2\pi}~.
\end{align}
This simple example shows that dimensional and cutoff regularizations might give rise to different 
on-shell $T$ matrices in a nonperturbative calculation. 
This is another instance of this issue (involving now an energy-dependent 
potential), which is  discussed in depth in Ref.~\cite{phillips.180319.1} for energy-independent potentials. 
 The differences in the results are shown in this reference to be due to causality  
(Wigner bound) that has a clear impact on cutoff regularization, but it is not so clear how it reflects on 
dimensional regularization (though these two methods agree in {\it perturbative} QFT calculations).

We can also shed light on the difficulties that dimensional regularization could have when applied 
in nonperturbative calculations by evaluating the compositeness $X$.
For that we need the half-off-shell $T$ matrix, which in dimensional regularization is
\begin{align}
\label{180324.11}
T(k,p)_{DR}&=\frac{(k^2+p^2)/(2k^2)}{\alpha+\frac{imk}{2\pi}}~,
\end{align}
from where we obtain for the coupling function squared
\begin{align}
\label{180324.12}
g_{DR}^2(p^2)&=\left(\frac{p^2-\gamma^2}{2\gamma^2}\right)^2\frac{2\pi\gamma}{m^2}~.
\end{align}
By applying Eq.~\eqref{170930.6} we can calculate straightforwardly 
 the compositeness in dimensional regularization $X_{DR}$ [it gives the same value for the integrals 
as Eq.~\eqref{171107.4}], with the result
\begin{align}
\label{180324.13}
X_{DR}&=\frac{1}{4\gamma^4}\left.\frac{\partial}{\partial k} k(k^2-\gamma^2)^2\right|_{k=i\gamma}=3~.
\end{align}
This is certainly a nonsense  because from general principles we know that $0\leq X \leq 1$. 
 This calculation then shows the potential problems of applying dimensional regularization to 
non-perturbative calculations. 

Let us now  calculate $X$ in an arbitrary type of cutoff regularization, $X_{\Lambda}$. 
 We have from Eq.~\eqref{180324.8} for the coupling function $g_\Lambda^2(p^2)$,
\begin{align}
\label{180324.14}
g_\Lambda^2(p^2)&=\left(1-(\gamma^2+p^2)\frac{\rho_\Lambda}{\Lambda^2}+{\cal{O}}(\Lambda^{-3})\right)
g_\Lambda^2(-\gamma^2)~,\\
g_\Lambda^2(-\gamma^2)&=\frac{2\pi/m^2\gamma }{1-\gamma r_s}~.\nn
\end{align}
In terms of it we have for $X_\Lambda$,
\begin{align}
\label{180324.15}
X_\Lambda&=g_\Lambda^2(-\gamma^2)2\left(\frac{m}{\pi}\right)^2
\int_0^\infty dp \frac{p^2}{(p^2+\gamma^2)^2}
\left(1-(\gamma^2+p^2)\frac{\rho_\Lambda}{\Lambda^2}+{\cal{O}}(\Lambda^{-3})\right)~.
\end{align}
Now, after regularizing the divergent integral one has that 
in the limit $\Lambda\to \infty$ the contribution to $X_\Lambda$ from terms 
suppressed by $\Lambda^{-2}$ and higher inverse power of $\Lambda$ do not contribute because 
the resulting integration is only linearly divergent in $\Lambda$. 
This is the slowest degree of vanishing because there are no higher powers 
of $p^2$ in the half-off-shell amplitude $T_\Lambda(k,p;E)$ 
[as we have worked out explicitly from the general solution from Eq.~\eqref{180319.8}].
 Thus, in the limit $\Lambda\to \infty$ we have
\begin{align}
\label{180324.16}
X_\Lambda&=\frac{1}{1-\gamma r_s}~.
\end{align}
This result, which is independent of the cut-off regularization method employed, 
gives $0\leq X_\Lambda \leq 1$ for $r_s\leq 0$. Working out the explicit expression of 
$\gamma$ as a function of $a_s$ and $r_s$ from Eq.~\eqref{180324.7} one has 
\begin{align}
\label{180708.1}
\gamma=\frac{1}{r_s}\left(1\pm \sqrt{1-\frac{2r_s}{a_s}}\right)~.
\end{align}
Thus, for $r_s\leq 0$ we only have the branch in Eq.~\eqref{180708.1} with the 
minus sign ($\gamma\geq 0$), which implies that $r_s/a_s\leq 0$ and then $a_s> 0$. 
The Eq.~\eqref{180324.16} simplifies to
\begin{align}
\label{180708.2}
X_\Lambda&=\frac{1}{\sqrt{1-2r_s/a_s}}\leq 1 ~,~r_s\leq 0~,~a_s>0 ~.
\end{align}
The issue of having a 
positive effective range when using cutoff regularization for an energy-independent 
potential (while requiring it to be Hermitian) \cite{phillips.180319.1}, as well as 
for the energy-dependent potential of Eq.~\eqref{180323.9b}, 
 has another manifestation here.  For $r_s>0$ the branch in Eq.~\eqref{180708.1}
 with the plus sign is the one possible for $a_s<0$, 
while the two branches of $\gamma$ are allowed for $a_s>0$ and $a_s/r_s\geq 2$.  
Despite that the potential in this case keeps its real character the 
compositeness becomes larger than 1, which is unacceptable. 
Therefore, having $r_s>0$ is not either compatible with the  
potential of Eq.~\eqref{180324.2}.

 In summary, a detailed analysis of the regularization and renormalization process is required for energy-dependent potentials 
between particle states in the continuum in order to conclude whether the result for $X$ is independent of 
the regularization method used. This is in contrast with the the general results for energy-independent potentials, 
Eqs.~\eqref{180320.8} and \eqref{180324.17b}, as well as for the general relation of Eq.~\eqref{180323.9}.
 We have studied the potential of Eq.~\eqref{180324.2} for which the on-shell $T$ matrix and $X$ are different between cutoff and 
dimensional regularization, with $X$ having an absurd value for the later. This is an extra deficiency of dimensional regularization when used 
in some nonperturbative calculations, in addition to those already analyzed in Ref.~\cite{phillips.180319.1} for energy-independent 
potentials. 
 However, in all the examples considered here the result for the compositeness is the same for any sort of cutoff regularization 
employed in the limit $\Lambda\to \infty$, similarly as happens for the on-shell $T$-matrix. 
As indicated, we are not able to provide a proof that this is always the case within a NR QFT calculation involving a singular 
potential, for which the calculation of physical results requires regularization and renormalization. 
 A general nonperturbative analysis is still lacking, though we think on physical grounds 
 that the compositeness of a bound state would come out as a derived 
quantity from the knowledge of the $S$-matrix, which should contain the spectroscopical information of the quantum 
mechanical problem. 
% This would generalize  the result
% that holds for regular potentials \cite{ref.170928.1} in a physically suitable way.

 %%%%%%%%%%%%%%%%%%%%%%%%%%%%%%%%%%%%%%%%%%%%%%%%%%%%%%%%%%%%%%%%%%%%%%
 \section{Relativistic bound state}
 \label{sec.170930.2}

 Up to the best of our knowledge there is no a general criterion for a relativistic bound state to be
 qualified as elementary. In the relativistic case one generally relies on the study of the wave-function
 renormalization and there is a series of results within specific models, like the Lee model
 \cite{amado.170930.1}
 or Yukawa type of interactions \cite{salam.170930.1,lurie.170930.1}.
 For these cases Refs.~\cite{amado.170930.1,salam.170930.1,lurie.170930.1}
 conclude that a bound state with $Z=0$ is purely composite.  
 Relativistic models with Yukawa-like interaction
 have been revisited frequently in the recent literature, e.g. in
 Refs.~\cite{hyodo.170930.1,nagahiro.180603.1,akaki.170930.1}.
 The property $0\leq Z\leq 1$ can be  obtained from the K\"allen-Lehmann representation,
 if  the interaction  Lagrangian does not involve field derivatives and the integral of
 the spectral function is finite, see e.g. Refs.~\cite{itzykson.171112.1,ref.170928.3}. 

 The straightforward extrapolation of the definition of $X$  in
 Eq.~\eqref{170929.13} cannot be given because contributions of eigenstates of $H_0$ belonging
 to the continuum spectrum  with different number of asymptotic particles can be
 generated by the standard conversion of energy into matter. In this way, Eq.~\eqref{170929.8}
 for the representation of $|\psi_B\rangle$ in terms of eigenstates of $H_0$ generalizes to
 \begin{align}
  \label{170930.12}
  |\psi_B\rangle &= \int d\gamma C_\gamma |AB_\gamma\rangle
       +\int d\eta D_\eta |AAB_\eta\rangle+\int d\mu \,\delta_\mu |ABB_\mu\rangle+\ldots\\
       &+\int d\eta_\nu F_\nu|CD_\nu\rangle+\ldots
       +\sum_n C_n|\varphi_n\rangle+\sum_n \int d\alpha C_{n\alpha} |A_\alpha\varphi_n\rangle
       +\ldots+\sum_{n,m} C_{nm} |\varphi_n\varphi_m\rangle+\ldots \nn
 \end{align}
 with quite an obvious notation.

 Nonetheless, we can
 still take advantage of the use of the number operators which are defined
 in the relativistic case as in NR QFT, cf. Eq.~\eqref{170929.9}. E.g. the average number of
 asymptotic particles of type $A$ in $|\psi_B\rangle$ as given by the decomposition in
 Eq.~\eqref{170930.12} is
 \begin{align}
   \label{170930.13}
   \langle \psi_B|N_D^A|\psi_B\rangle&=\int d\gamma |C_\gamma|^2 
   +2\int d\eta |D_\eta|^2 +    \int d\mu |\delta_\mu|^2+\ldots+\sum_n \int d\alpha
   |C_{n\alpha}|^2+\ldots   
 \end{align}
 In this way we can deduce the following universal criterion for a bound state
 to be considered as elementary with respect to the particles in the continuum spectrum, 
 applicable both in the relativistic and NR cases:
 \begin{align}
   \label{170930.14}
   \langle \psi_B|N_D^A|\psi_B\rangle&=0~~,~~\forall A~.
 \end{align}
 %\footnote{  Of course, at the practical level we can consider that a bound state is
% approximately elementary if 
%\begin{align}
%   \label{170930.14b}
%   \langle \psi_B|N_D^A|\psi_B\rangle&\ll 1~~,~~\forall A~.
%\end{align}}
 Strictly speaking we have another extra condition in addition
 to Eq.~\eqref{170930.14} for relativistic systems in order 
to avoid the possibility that $|\psi_B\rangle$ had components  of states made by several bare
 elementary discrete states, as the last contribution shown in Eq.~\eqref{170930.12}. Denoting by $N_D^E$ the
sum of the number operators for the bare elementary discrete states ($N_D^n$), $N_D^E=\sum_n N_D^n$, one also
has to discard that
\begin{align}
  \label{171002.2}
  \langle \psi_B|N_D^E|\psi_B\rangle > 1~.
  \end{align}

Another consequence that can be extracted by evaluating the expectation value of the
 number operators  in $|\psi_B\rangle $, Eq.~\eqref{170930.13},
is  the following. Let us consider that for a particle species $A$ one has that  
\begin{align}
  \label{170930.15}
  \langle \psi_B|N_D^A|\psi_B\rangle=x_A
\end{align}
with $x_A\geq m$ and $m\geq 0$ a natural number. In such circumstances we can conclude that  the
 free-particle states containing $m$ or more  asymptotic particles of type $A$ are relevant
in the bound state $|\psi_B\rangle$. 

%%%%%%%%%%%%%%%%%%%%%%%%%%%%%%%%%%%%%%%%%%%%%%%%%%%%%%%%%%%%%%%%%%%%%%%%%%%%%%%%%%
\section{Calculation of $\langle \psi_B|N_D^A|\psi_B\rangle $ in relativistic QFT}
\label{sec.170930.3}

Let us now discuss the calculation of the expectation value $\langle \psi_B|N_D^A|\psi_B\rangle $ in
relativistic QFT. We follow the same steps as introduced in Sec.~\ref{sec.170929.2} for NR QFT,
since many of them are equally valid in the relativistic case.
 Being specific,  Eqs.~\eqref{170929.14}, \eqref{170929.15} can be used also now 
and then instead of Eq.~\eqref{170930.1b} for $X$ we have the analogous expression
\begin{align}
  \label{170930.16}
  \langle \psi_B|N_D^A| \psi_B\rangle &=
  \langle \varphi_B|U(+\infty,0)N_D^A U(0,-\infty)|\varphi_B\rangle~.
  \end{align}
Again the extra time evolution from 0 to $t$ in Eqs.~\eqref{170930.2} and \eqref{180603.3}
can be equally applied in the relativistic case (of course, here also $[N_D^A,H_0]=0$ \cite{ref.170928.4}).
We then arrive to the time-ordered expression for $\langle \psi_B|N_D^A|\psi_B\rangle$ ready to be applied
in QFT:
\begin{align}
  \label{170930.17}
  \langle \psi_B|N_D^A|\psi_B\rangle&=  \lim_{T\to +\infty}\frac{1}{T}\int_{-T/2}^{+T/2} dt
  \langle \varphi_B|U_D(+\infty,t)  N_D^A(t)U_D(t,-\infty)|\varphi_B\rangle~.
\end{align}
The infinite factor $T$ in the denominator of this equation cancels with the Dirac delta function of total
energy conservation.

We can also express the number operator $N_D^A$ in terms of free fields, analogously as done in
the non-relativistic case. Let us a consider a scalar particle $A$ and  define the free fields
\begin{align}
  \label{171110.4}
  \psi^{(+)}(x)&=\int\frac{d^3\vq}{(2\pi)^3}a(\vq)e^{-i\tilde{q} x}~,\\
  \psi^{(-)}(x)&=\int\frac{d^3\vq}{(2\pi)^3}a^\dagger(\vq) e^{i\tilde{q} x}=\psi^{(+)}(x)^\dagger~,\nn
\end{align}
so that
\begin{align}
  \label{171110.5}
  N_D^A(t)&=-2i\int d^3 x \dot{\psi}^{(-)}(x)\psi^{(+)}(x)~.
\end{align}
with $\dot{\psi}^{(-)}(x)=\frac{\partial \psi^{(-)}(x)}{\partial t}$. 
 The following expression for the expectation value $\langle \psi_B|N_D^A|\psi_B\rangle$ results
\begin{align}
  \label{171110.6}
\langle \psi_B|N_D^A|\psi_B\rangle   &=-2i\lim_{T\to +\infty}\frac{1}{T}\int d^4x 
\langle \varphi_B| P\left[ e^{-i\int d^4 x' {\cal H}_D(x')}
  \dot{\psi}^{(-)}(x)\psi^{(+)}(x) \right] |\varphi_B\rangle~.
\end{align}
where the interaction have been written in terms of an interaction-Hamiltonian density
${\cal H}_D(x)$ in the Dirac picture. 

\begin{figure}
\begin{center}
\includegraphics[width=0.25\textwidth]{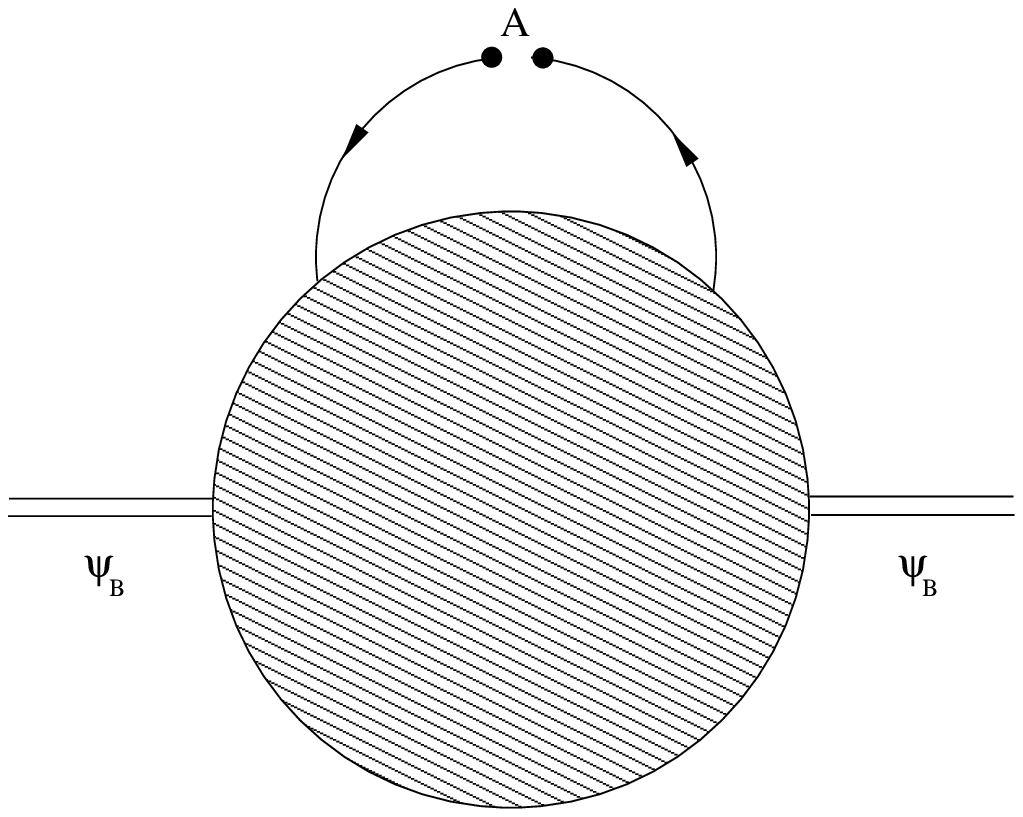}
\caption{Schematic representation of the Feynman diagrams for the calculation of
  $\langle \psi_B|N_D^A|\psi_B\rangle$ in QFT, Eqs.~\eqref{170930.17} or \eqref{171110.6}.
  The insertion of the number operator for
the particle species $A$  is indicated by the double dot.}
\label{fig.171110.1}
\end{center}
\end{figure}

The evaluation of Eqs.~\eqref{170930.17} or \eqref{171110.6}
involves (infinite) more diagrams than in NR QFT. The set of Feynman diagrams involved
can be schematically represented as in Fig.~\ref{fig.171110.1}, where the shaded circle
represents any set of connected vertices without any insertion of the number operator which is
indicated by the double dot.
 However, in order to apply Feynman rules to the calculation of Eq.~\eqref{171110.6}
one has to take into account that the two internal lines in Fig.~\ref{fig.171110.1}
ending in any of the double dots is not a standard Feynman propagator.
 According to the Wick theorem for a neutral scalar field they correspond to
\begin{align}
\label{171110.8}
  \langle \varphi_0|P\left[\psi(x_1)\,\psi^{(+)}(x_2) \right]|\varphi_0\rangle&=
  i\int \frac{d^4 k}{(2\pi)^4}\frac{e^{i k (x_1-x_2)}}{2E_k(k^0-E_k+i\vep)}~,\\
\label{171110.9}
  \langle \varphi_0|P\left[\psi(x_1)\,\psi^{(-)}(x_2) \right]|\varphi_0\rangle&=
  i\int \frac{d^4 k}{(2\pi)^4}\frac{e^{i k (x_2-x_1)}}{2E_k(k^0-E_k+i\vep)}~,\\
%\label{171110.7}
  \psi(x_1)&=\psi^{(+)}(x_1)+\psi^{(-)}(x_1)~.\nn
  \end{align}
Here, $|\varphi_0\rangle$ is the non-interacting vacuum and $E_k=\sqrt{m_A^2+k^2}$, 
with $m_A$ the physical mass. As a result,
any of the two internal lines explicitly shown in Fig.~\ref{fig.171110.1} correspond to
\begin{align}
  \label{171110.10}
  \frac{i}{2E_k(k^0-E_k+i\vep)}~,
\end{align}
instead of a standard Feynman propagator for a scalar field. 
The arrows in the same lines in Fig.~\ref{fig.171110.1} refer to the momentum flow according
to Eqs.~\eqref{171110.8} and \eqref{171110.9},
 such that a line ends at $\psi^{(+)}(x)$ and another one leaves at $\psi^{(-)}(x)$.

% The Eq.~\eqref{171110.6} can be also written straightforwardly in terms of a three-point Green function
%by including an interpolation field for the bound state $|\psi_B\rangle$ and then applying
%standard reduction techniques in QFT \cite{ref.170928.3,itzykson.171112.1}.
% For conciseness we skip any further detail on this here, though
% it seems to be indeed the most appropriate way to proceed for a relativistic QFT calculation.

%Another way is to consider the scattering between (two) particles in the presence of the
%number operator density  and then extract the residue of the scattering amplitude
%at the double bound-state pole.
% For conciseness we skip any further detail on this here.

Two limit cases are worth pointing out.
 In the case in which the bound state occurs nearby a two-body threshold the new formalism reduces
 to the NR case again.\footnote{For a relativistic version of the integral equation for the coupling
   of the bound-state with the continuum asymptotic states the reader is referred to Ref.~\cite{salam.170930.1}, 
   which employs the Bethe-Salpeter equation instead of the LS equation.}
 Furthermore, if one can conclude that only two-body channels  dominate
 then one expects that the more important Feynman diagrams are those of Fig.~\ref{fig.170930.1}.

 For a given total Hamiltonian $H$ the expectation values of $N_D^A$ are invariant under unitary transformations and field
reparametrizations. This is a direct consequence of Eq.~\eqref{170930.13}. Nonetheless, its evaluation is nonperturbative and it 
is beyond the scope of the present manuscript to study the possible regulator independence of $X$ (beyond the discussions  
given above for NR QFT, cf. Sec.~\ref{sec.170930.1}).

% It is worth stressing that for a given total Hamiltonian $H$ the expectation values $N_D^A$ are
% observable in the sense that they are invariant under unitary transformations and field
%reparametrizations. This is a direct consequence of Eq.~\eqref{170930.13}.

 %%%%%%%%%%%%%%%%%%%%%%%%%%%%%%%%%%%%%%%%%%%%%%%%%%%%%%%%%%%%%%%%%%%%%%%%%%%%%%%%%%%%%%%%%%%%%%%%%%%%%%%%%%%
 \section{Resonances}
 \label{sec.170930.4}

 In this section we discuss the generalization of many of the results given in Secs.~\ref{ref.170929.1}--\ref{sec.170930.3}
 to the case of resonance states. The latter correspond to poles of the $T$ matrix in an unphysical RS that can be
 reached by the analytical extrapolation, typically in the complex energy or $s$ plane for
 NR and relativistic cases, respectively (with $s$ the usual Mandelstam variable). We assume in the following that 
 the pole is of order one.
 
 %%%%%%%%%%%%%%%%%%%%%%%%%%%%%%%%%%%%%%%%%%%%%%%%%%%%%%%%%%%%%%%%%%%%%%%%%%%%%
 \subsection{Definitions and QFT formalism}
 \label{sec.170930.5}

 An approximate way to afford the problem of evaluating $Z$ in the non-relativistic case for an unstable particle
 near a two-body threshold was considered in Refs.~\cite{markusin.170930.2,baru.170930.1}.
 The approach is based on integrating the spectral density
 of a bare elementary discrete state around the resonance signal region, 
 in such a way that if this integral is small the state is mostly composite while if
 it is close to 1 then it is mostly elementary. These results have also a clear connection with the
 counting pole rule of Morgan \cite{morgan.170930.1} and with the possible presence of near Castillejo-Dalitz-Dyson
 poles \cite{kang.170930.1}.

 Let us continue here with our interpretation of the compositeness $X$
 based on the definition of Eq.~\eqref{170929.13} in NRQM.
 We derive our results for resonance states by the analytical continuation
 of the expressions from the physical energy axis. In this respect, let us first discuss
 which is the matrix element that one should extrapolate analytically in order to reach the
 resonance pole.

 The most straightforward option would be to calculate the expectation value
 of the operator number $N_D$ in an in state, $|\psi^+_\alpha\rangle$.
 For definiteness let us take a two-body in state of
 particles $A$ and $B$. In the same way that it is demonstrated that
 $\langle \psi_\alpha^+|\psi_\alpha^+\rangle=\langle \varphi_\alpha|\varphi_\alpha\rangle$
 \cite{ref.170928.3} one concludes that 
 \begin{align}
  \label{170930.18}
   \langle \psi_\alpha^+|N_D^A+N_D^B|\psi_\alpha^+\rangle=2\langle \varphi_\alpha|\varphi_\alpha\rangle~.  
 \end{align}
 However, this matrix element cannot be analytically continued to the resonance pole
 at $E_R=M_R-i\Gamma/2$ in the 2nd RS. The reason is
 because of the bra $\langle \psi_\alpha^+|$, which obeys the equation (it can be derived from
 $|\psi^+_\alpha\rangle=U(0,-\infty)|\varphi_\alpha\rangle$, see e.g. Ref.~\cite{ref.170928.3}) 
 \begin{align}
 \label{170930.20}
 \langle \psi_\alpha^+ | &=\langle \varphi_\alpha|
 +\int d \gamma \frac{T_{\alpha\gamma}(E_\alpha+i\vep)^\dagger}{E_\alpha-i\vep-E_\gamma}\langle \varphi_\gamma|
 +\sum_n \frac{T_{\alpha n}(E_\alpha+i\vep)^\dagger}{E_\alpha-i\vep-E_n}\langle \varphi_n|\nn\\
 &=\langle \varphi_\alpha|
 +\int d \gamma \frac{T_{\alpha\gamma}(E_\alpha-i\vep)}{E_\alpha-i\vep-E_\gamma}\langle \varphi_\gamma|
 +\sum_n \frac{T_{\alpha n}(E_\alpha-i\vep)}{E_\alpha-i\vep-E_n}\langle \varphi_n|~.
 \end{align}
 Here we have taken into account that $T(E\pm i\vep)^\dagger=T(E\mp i\vep)$ as follows from the
 LS equation, Eq.~\eqref{170930.7}. The analytically continuation of Eq.~\eqref{170930.20}
 to $E=E_R$ must be done in the 1st RS because the imaginary part of the
 energy is already negative. 

\begin{figure}
\begin{center}
\includegraphics[width=0.3\textwidth]{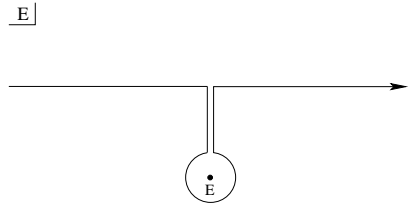}
\caption{Deformation of the integration contour along the physical energy in Eq.~\eqref{171001.2}
 needed to reach the resonance pole at $E_R=M_R-i\Gamma/2$.}
\label{fig.171001.1}
\end{center}
\end{figure}

 In order to reach the resonance pole we have to use the bra of an out state and the ket
 of an in state, as it is the case when evaluating the $S$-matrix elements. 
 For the bra of the out state instead of Eq.~\eqref{170930.20} we have 
\begin{align}
 \label{171001.2}
 \langle \psi_\alpha^- | &=\langle \varphi_\alpha|
 +\int d \gamma \frac{T_{\alpha\gamma}(E_\alpha-i\vep)^\dagger}{E_\alpha+i\vep-E_\gamma}\langle \varphi_\gamma|
 +\sum_n \frac{T_{\alpha n}(E_\alpha-i\vep)^\dagger}{E_\alpha+i\vep-E_n}\langle \varphi_n|\nn\\
 &=\langle \varphi_\alpha|
 +\int d \gamma \frac{T_{\alpha\gamma}(E_\alpha+i\vep)}{E_\alpha+i\vep-E_\gamma}\langle \varphi_\gamma|
 +\sum_n \frac{T_{\alpha n}(E_\alpha+i\vep)}{E_\alpha+i\vep-E_n}\langle \varphi_n|~,
 \end{align}
and now its analytical extrapolation to $E=E_R$ requires to cross the unitarity cut ($E>0$) to reach
energy values with  ${\rm Im}E<0$, and then one moves to the 2nd RS.
The analytical continuation of the previous equation requires
to deform the integration contour along the physical axis of energy as shown in Fig.~\ref{fig.171001.1}.

In this way,
 instead of the expectation value of Eq.~\eqref{170930.18} one should consider the matrix
 element
 \begin{align}
   \label{171001.1}
 \langle \psi^-_\alpha|N_D|\psi^+_\alpha\rangle~,
 \end{align}
 with $N_D$ defined in Eq.~\eqref{170930.1}, and extrapolate it to the resonance pole position.
 The previous matrix element has a double pole at the resonance pole position, because of the initial and final
 state interactions. The residue of this double pole divided by the coupling
 squared is the expectation value of the operator $N_D$ in the resonance states  \cite{albaladejo.171001.1}.

 This limit process can be avoided if we use an analogous formalism to
 that explained in Sec.~\ref{sec.170929.2},
 but now for resonance states. We express the in/out resonance state $|\psi^\pm_R\rangle$ by evolving the
 bare one $|\varphi_R\rangle$ from asymptotic times\footnote{To talk about the bare resonance state might be
   qualified as an abuse of language. Nonetheless, it is consistent since $H_0|\varphi_R\rangle=E_R|\varphi_R\rangle$
   as follows by relating $|\psi_R^\pm\rangle$ with $|\varphi_R\rangle$ with analogous equations to
   Eqs.~\eqref{170930.20} and \eqref{171001.2}. Additionally one ends with the same results as obtained by
   proceeding with the analytical continuation of Eq.~\eqref{171001.1} as already discussed.}

 \begin{align}
  \label{171001.3}
  |\psi^+_R\rangle &=U_D(0,-\infty)|\varphi_R\rangle~,\\
  \langle \psi_R^-| &= \langle \varphi_R|U_D(+\infty,0)~.\nn
\end{align}
 Thus,
 \begin{align}
  \label{171001.4}
  X&=\frac{1}{n}\langle \varphi_R|U(+\infty,0)N_D U(0,-\infty)|\varphi_R\rangle~,
\end{align}
Next, we introduce the  extra time evolution from 0 to $t$. For that let us notice
that
\begin{align}
  \label{171001.5}
  U_D(t,0)|\psi_R^+\rangle&=e^{iH_0 t} e^{-i Ht}|\psi_R^+\rangle=e^{-(iM_R+\frac{\Gamma}{2})t}e^{iH_0 t}U_D(0,-\infty)
  |\varphi_R\rangle~,\\
  \langle \psi^-_R|U_D(0,t)&= \langle \psi^-_R|e^{iHt}e^{-iH_0t}=\langle \varphi_R|U_D(+\infty,0)e^{-iH_0 t}
  e^{(iM_R+\frac{\Gamma}{2})t}~.\nn
\end{align}
The  factors $e^{\pm(iM_R+\frac{\Gamma}{2})t}$  cancel between them  in Eq.~\eqref{171001.4} while
$e^{iH_0t}N_De^{-iH_0t}=N_D(t)=N_D$. As a result we can write 
\begin{align}
    \label{171001.6}
    X=\frac{1}{n}\lim_{T\to +\infty}\frac{1}{T}\int_{-T/2}^{+T/2} dt \langle \varphi_R|U_D(+\infty,t)
    N_D(t)U_D(t,-\infty)|\varphi_R\rangle~,
\end{align}

  The same steps as in Eqs.~\eqref{171110.1}--\eqref{171110.3} allow us to re-express $N_D$ in
terms of NR fields $\psi_i(x)$.
Therefore, for scalar particles $A_i$ we end with the following expression for $X$ 
\begin{align}
  \label{171110.11}
X &=\frac{1}{n}\lim_{T\to +\infty}\frac{1}{T}\int d^4x 
\langle \varphi_R| P\left[ e^{-i\int_{-\infty}^{+\infty}dt' V_D(t')}
 \sum_i \psi_{A_i}^\dagger(x)\psi_{A_i}(x)
 \right] |\varphi_R\rangle~.  
\end{align}
analogous to Eq.~\eqref{171110.3}.

In the case of two-particle asymptotic states the calculation of $X$ can be done by evaluating the
Feynman diagrams of Fig.~\ref{fig.170930.1}. Performing the corresponding partial-wave decomposition
as in Eq.~\eqref{170930.5} one has the following expression for $X_{\ell S}$
\begin{align}
  \label{171001.7}
  X_{\ell S}&=\frac{1}{2\pi^2}\int_0^\infty dk k^2\frac{g_{\ell S}^2(k^2)}{(k^2/2\mu-E_R)^2}
  +\frac{i\mu^2}{\pi\varkappa}\left.\frac{\partial kg_{\ell S}^2(k^2)}{\partial k}\right|_{k=\varkappa}~.
  \end{align}
Compared with Eq.~\eqref{170930.6} there is an extra term due to the deformation of the analytical contour
for integration, as shown in Fig.~\ref{fig.171001.1}.
 Because of the same reason, the homogeneous integral
equation satisfied by $g(k)$ (a matrix notation should be used if several partial waves mix) is now
\begin{align}
  \label{171001.8}
%  g(k)&=\frac{\mu}{\pi^2}\int_0^\infty dk'{k'}^2\frac{V(k,k')g(k')}{\varkappa^2-{k'}^2}
%  -\frac{i\mu \varkappa V(k,\varkappa)/\pi}{1+i\mu \varkappa V(\varkappa,\varkappa)/\pi}
%  \,\frac{\mu}{\pi^2}\int_0^\infty dk'{k'}^2\frac{V(\varkappa,k')g(k')}{\varkappa^2-{k'}^2}~.
  g(k)&=\frac{\mu}{\pi^2}\int_0^\infty dk'\frac{{k'}^2 g(k')}{\varkappa^2-{k'}^2}
  \left[V(k,k')
  -\frac{i\mu \varkappa V(k,\varkappa)V(\varkappa,k')/\pi}{1+i\mu \varkappa V(\varkappa,\varkappa)/\pi}\right]~.
\end{align}
From this equation it is also clear that $g_{\ell S}(-k)=(-1)^{\ell}g_{\ell S}(k)$ and
then $g_{\ell S}^2$ is a function of $k^2$, as reflected in Eq.~\eqref{171001.7}. 
We can also determine the coupling function $g(k)$ by calculating the residue of the 
(half-)off-shell $T$ matrix at the resonance pole position, 
  \begin{align}
  \label{180327.1}
  g(k)g(p)&=\lim_{E\to E_R}(E-E_R)T(k,p;E)~.
  \end{align}
At the resonance pole position the on-shell three-momentum is denoted by 
$\varkappa$ which is defined in the 2nd RS as $\varkappa=\sqrt[II]{2\mu E_R}$, 
where $\sqrt[II]{z}$ is defined in the 2nd RS with ${\rm arg}z\in [2\pi,4\pi[$. 
%]$ 

     Although Eq.~\eqref{171001.7} is not explicitly real and positive, we can show that $X=1$ 
for an energy-independent potential. 
We follow analogous steps as performed in Sec.~\ref{sec.180319.1} for a zero-range potential and in Sec.~\ref{sec.180321.1} 
for a regular or singular finite-range potential. 
In both cases the only change concerns the analytical extrapolation of the 
full-off-shell $T$-matrix, cf. Eq.~\eqref{180319.5}, from the 1st to the 2nd RS by taking into account the deformation of the integration contour 
of Fig.~\ref{fig.171001.1}.
 This analytical extrapolation only affects the matrices $[G_\alpha(E)]$. 
We give explicit expressions for the zero-range potential case, since the same expressions are valid 
for finite-range potentials under the exchange $[q_\alpha]\to [f_\alpha(q_\alpha)]$, once they are approximated with arbitrary precision 
by a separable potential of rank $N$, Eq.~\eqref{180321.5} (for singular potentials this would require regularization, as discussed 
in Sec.~\ref{sec.180321.1}).

Instead of $[G_\alpha(E)]$ we have now its analytical extrapolation in the 2nd RS, $[G^{II}_\alpha(E)]$, given by
 \begin{align}
\label{180328.1}
[G^{II}_\alpha(E)]&=\frac{m_\alpha}{\pi^2}\int_0^\infty dq\frac{q^2}{q^2-2m_\alpha E}[q_\alpha]\cdot [q_\alpha]^T
+\frac{im_\alpha}{\pi}\sqrt[II]{2m_\alpha E}\left[\sqrt[II]{2m_\alpha E}\right]\cdot \left[\sqrt[II]{2m_\alpha E}\right]^T~.
 \end{align}    
Then, the $T$ matrix in the 2nd RS
 \begin{align}
   \label{180328.2}
 t^{II}_{\alpha\beta}(k_\alpha,p_\beta;E)&=[k_\alpha]^T\cdot [t^{II}(E)]\cdot [p_\beta]~,
 \end{align}
  can be calculated making use of an analogous equation to Eq.~\eqref{180319.8},
\begin{align}
   \label{180328.3}
         [t^{II}(E)]&=[D^{II}(E)]^{-1}~,\\
         [D^{II}(E)]&=[v(E)]^{-1}+[G^{II}(E)]~.
\end{align}
In terms of it we can calculate the coupling functions from the residue of the $T$ matrix at the resonance pole position
 \begin{align}
\label{180328.4}
   g_\alpha(k_\alpha)g_\beta(p_\beta)&=\lim_{E\to E_R}(E-E_R)t^{II}_{\alpha\beta}(k_\alpha,p_\beta;E)
   =\left.\frac{[k_\alpha]^T\cdot [d^{II}]\cdot [p_\beta]}{{\Delta^{II}}'}\right|_{E=E_R,p_\beta=\varkappa_\beta}~,\\
   {\Delta^{II}}'&=\left.\frac{\partial \Delta^{II}(E)}{\partial E}\right|_{E=E_R}~,\nn
 \end{align}
where $\Delta^{II}(E)$ is the determinant of $[D^{II}(E)]$, $[d^{II}(E)]$ is its adjoint matrix 
 and $\varkappa_\beta=\sqrt[II]{2m_\beta E_R}$. 
An analogous formula for $g_\alpha^2(k_\alpha)^2$ to that already obtained in the bound-state case results
\begin{align}
   \label{180328.5}
 g^2_\alpha(k_\alpha^2)&=\left.\frac{([k_\alpha]^T\cdot [d^{II}] \cdot [p_\alpha])^2}{[p_\alpha]^T\cdot
   [d^{II}]\cdot \frac{\partial [D^{II}]}{\partial E}\cdot [d^{II}]\cdot [p_\alpha]}\right|_{E=E_R,p_{\alpha}=\varkappa_{\alpha}}~.
 \end{align}

The compositeness $X_\alpha$ for the resonance state then reads, cf. Eq.~\eqref{171001.7},
\begin{align}
   \label{180328.6}
X_\alpha&=\left([p_\alpha]^T\cdot [d^{II}]\cdot \frac{\partial[D^{II}]}{\partial E}\cdot [d^{II}]\cdot [p_\alpha]\right)^{-1}\\
&\times  \frac{\partial}{\partial E}\left.\left(
\frac{m_\alpha}{\pi^2}\int_0^\infty dk\frac{k^2}{k^2-2m_\alpha E} [k_\alpha]^T\cdot [d^{II}]\cdot [p_\alpha]\, [k_\alpha]^T\cdot [d^{II}]\cdot [p_\alpha]
\right.\right. \nn\\
&\left.\left.+i
\frac{m_\alpha}{\pi}\sqrt[II]{2m_\alpha E}\,[\sqrt[II]{2m_\alpha E}]^T\cdot [d^{II}]\cdot [p_\alpha]\,[\sqrt[II]{2m_\alpha E}]^T\cdot [d^{II}]\cdot [p_\alpha]
\right)\right|_{E=E_R,p_\alpha=\varkappa_\alpha}~.\nn
\end{align}
For an energy-independent potential $\partial [D^{II}]/\partial E=\partial [G^{II}]/\partial E$. 
Following them completely analogous steps as in Eqs.~\eqref{180320.7}--\eqref{180320.8} 
 we have also for a resonance that the total compositeness is 1, 
\begin{align}
   \label{180328.7}
\sum_{\alpha=1}^n X_\alpha&=1~.
\end{align}
 
This result is valid as well for a finite-range energy-independent singular or regular  potential
 by following the same steps as 
in Sec.~\ref{sec.180321.1}, replacing $[G(E)]$ by $[G^{II}(E)]$, with
 \begin{align}
\label{180328.1b}
[G^{II}_\alpha(E)]&=\frac{m_\alpha}{\pi^2}\int_0^\infty dq\frac{q^2}{q^2-2m_\alpha E}[f_\alpha(q)]\cdot [f_\alpha(q)]^T
+\frac{im_\alpha}{\pi}\sqrt[II]{2m_\alpha E}\left[f_\alpha(\sqrt[II]{2m_\alpha E})\right]\cdot 
\left[f_\alpha(\sqrt[II]{2m_\alpha E})\right]^T~.
 \end{align}    
 To avoid being too repetitive we refrain from reproducing them explicitly. 
The same conclusion was deduced in  Ref.~\cite{ref.171001.2}  for
 an energy-independent regular potential making use of the analytical extrapolation of the Schr\"odinger equation 
in the 2nd RS of the complex energy plane. However, our demonstration (based on the use of the LS equation) allows to treat singular potentials too. 
In addition, the normalization to one of the resonance state within our formalism is a consequence with a clear 
physical picture behind (the number of asymptotic particles in the state is 2), while in Ref.~\cite{ref.171001.2} 
it relies on a pure mathematical basis. Our conclusion that the total compositeness is 1 for a 
finite-range energy-independent potential implies that a resonance is then a purely composite state. 
 However, the compositeness $X$ is in general
a complex number for $\partial v(E)/\partial E\neq 0$ and
 we discuss below in Sec.~\ref{sec.170930.6} how one can still give sense to $X$.

As in Sec.~\ref{sec.180319.1} for a bound state, it is illustrative to write a closed formula for $X$ in the case of a resonance
 with an energy-independent zero-range potential, for which $X=1$ as just shown above. 
 We again employ the regularization method based on
 including a convergent factor in Eq.~\eqref{171001.7} (which gives the same results for the integrals 
as dimensional regularization in three dimensions).
 Since the resonance pole lies in the 2nd RS, with ${\rm Im}k<0$,
 now we  close the integration contour along the lower half plane in the $k$-complex plane.
 In this way, instead of Eq.~\eqref{171007.1} we now regularize the potential as
\begin{align}
  \label{171110.12}
  V(k',k)\to V(k',k) e^{-i\epsilon (k+k')}~,
\end{align}
including an extra minus sign in the exponent of the convergent factor.
 It is now straightforward to obtain
\begin{align}
  \label{171001.9}
  X_{\ell S}&=\frac{1}{4\pi^2}\int_{-\infty}^\infty dk k^2\frac{g_{\ell S}^2(k^2)e^{-i\vep k}}{(k^2/2\mu-E_R)^2}
  +\frac{i\mu^2}{\pi\varkappa}\left.\frac{\partial kg_{\ell S}^2(k^2)}{\partial k}\right|_{k=\varkappa}
 \nn \\
  &=g^2(\varkappa^2)\frac{i\mu^2}{2\pi\varkappa}
  +\frac{i\mu^2\varkappa}{\pi}\left.\frac{\partial g^2(k^2)}{\partial k^2}\right|_{k=\varkappa}~.
\end{align}
The first term is already well-known while the latter is a new contribution. 
By direct computation  Eq.~\eqref{171001.9} can also be expressed as
\begin{align}
  \label{171001.10}
  X_{\ell S}&=\frac{2\mu^2}{\pi^2}\int_0^\infty
 dk^2 \sqrt[{II}]{k^2+i\vep}\frac{g_{\ell S}^2(k^2)}{(k^2-\varkappa^2)^2}~.
\end{align}
%where $\sqrt[II]{z}$ is $\sqrt{z}$ in the 2nd RS with ${\rm arg}z\in [2\pi,4\pi[$. 
Notice that $\sqrt[II]{k^2+i\vep}=-\sqrt[II]{k^2-i\vep}=-k$.
	This equation is entirely equivalent to that for a bound state, Eq.~\eqref{170930.6}, but written in the 2nd RS 
as corresponds to a resonance state. It is also interesting to realize about the presence  
of the factor $g_{\ell S}^2(k^2)/(k^2-\varkappa^2)^2$ and not of its modulus, as it corresponds to a 
Gamow state \cite{ref.171001.2}.

For the relativistic case we can evaluate the matrix elements of the operator numbers
$N_D^A$ between resonances states. As in the bound state case we can directly export the
equations derived within NR QFT and use them also for relativistic QFT. Namely,
we are referring to Eqs.~\eqref{171001.3} and \eqref{171001.5}. In this way, we can write
the matrix element of $N_D^A$ between  in/out  resonance states in relativistic
QFT as
\begin{align}
  \label{171001.14}
\langle \psi_R^-|N_D^A|\psi_R^+\rangle&=\lim_{T\to +\infty}\frac{1}{T}\int_{-T/2}^{+T/2} dt \langle \varphi_R|U_D(+\infty,t)
    N_D^A(t)U_D(t,-\infty)|\varphi_R\rangle~. 
  \end{align}
As in the relativistic bound state case we can express the number operators $N_D^A(t)$ as bilinear operators of relativistic
fields. For $A$ being a scalar field we can use the result of Eq.~\eqref{171110.5} and write the matrix element of 
 $N_D^A$ between resonance states as
\begin{align}
  \label{171111.1}
\langle \psi_R^{(-)}|N_D^A|\psi_R^{(+)}\rangle   &=-2i\lim_{T\to +\infty}\frac{1}{T}\int d^4x 
\langle \varphi_R| P\left[ e^{-i\int d^4 x' {\cal H}_D(x')}
  \dot{\psi}^{(-)}(x)\psi^{(+)}(x) \right] |\varphi_R\rangle~.
\end{align}
 The set of Feynman diagrams is represented in Fig.~\ref{fig.171110.1}, with the obvious replacement
of $|\psi_B\rangle$ by $|\psi_R^{\pm}\rangle$ to the right and left, respectively.
Together with Eq.~\eqref{171111.1} one also has to keep in  mind the meaning of the internal lines
joining the field bilinear associated to the number operator, as explained in Sec.~\ref{sec.170930.3}, cf.
Eqs.~\eqref{171110.8}, \eqref{171110.9} and \eqref{171110.10}.
 For other particle species the expression of $N_D^A$ in terms of a field bilinear can be worked out straightforwardly. 
 Another convenient way to proceed is to re-express the right hand side of  Eq.~\eqref{171111.1}  
by including an interpolation field for the resonance state  and then applying
standard reduction techniques in QFT \cite{albaladejo.171001.1}.
Another possibility is to consider the scattering between particles in the presence of the
number operator density  and then extract the residue of the scattering amplitude
at the double resonance pole. This idea was applied to calculate the scalar form factor
of the $f_0(500)$ in Ref.~\cite{albaladejo.171001.1}
by evaluating $\pi\pi$ scattering in the presence of a scalar source.

It is obvious that a necessary condition   for a resonance being elementary is the 
 that the expectation value of the number operators of the asymptotic free particles of any 
 species be zero,
\begin{align}
  \label{171002.1}
  \langle \psi_R^-|N_D^A|\psi_R^+\rangle&=0 ~~,~~\forall A~.
\end{align}
In practical application it would be enough that
$|\langle \psi_R^-|N_D^A|\psi_R^+\rangle|\ll 1~,$ $\forall A$.
  As a clarification remark why we cannot state it as a sufficient condition as well,
 let us consider a decomposition of a resonance state
 as in Eq.~\eqref{170930.12} (with $|\psi_B\rangle$ replaced by $|\psi_R^+\rangle$).
 Now, by taking
 the expectation value in Eq.~\eqref{171002.1} we would only pick up contributions from those basis states
 including free particles of type $A$. However, as follows from the NR QFT  analysis for a resonance,
 cf. Eq.~\eqref{171001.10}, one should not expect
 to have the sum of the modules squared of the coefficients in the linear decomposition
 (as in Eq.~\eqref{170930.12} for a bound state) but rather the coefficients squared (at least for those channels
 that are open at the resonance mass) because of the analytical extrapolation to the resonance pole in the 2nd RS.
 Therefore, we have in general the sum of several complex numbers which can be zero even though they are not
 separately.

%%%%%%%%%%%%%%%%%%%%%%%%%%%%%%%%%%%%%%%%%%%%%%%%%%%%%%%%%%%%%%%%%%%%%%%%%%%%%%%%%%%%%%%%%%
\subsection{Phase-factor transformations}
 \label{sec.170930.6}

 The main point of Ref.~\cite{guo.170930.1} is to establish the existence of transformations
 at the level of the partial-wave projected $S$ matrix such that
 \begin{align}
  \label{171001.15}
  &S \to {\cal O}S{\cal O}^T\\
  &{\cal O}{\cal O}^\dagger=I~.
 \end{align}
 In order not to modify the modulus of the residues at the resonance pole the unitary matrix ${\cal O}$ 
 is taken diagonal. This can be probed to be the case for a narrow
 resonance\footnote{A resonance lying above
   threshold but with vanishing width. The width of its signal is fully reflected in the physical energy axis.}
   by invoking unitarity 
 and the physical requirement that the module of every coupling to an open channel properly determines
 its branching decay ratio or partial width, so that it should not be modified.\footnote{This also applies 
 to closed channels whose thresholds are much closer to the resonance mass than the width of the 
 resonance \cite{guo.170930.1}, e.g. the $f_0(980)$ and the $K\bar{K}$ channel.} However, its phase
 is quite arbitrary and it is determined by the smooth non-resonant contributions.
 For more details see Ref.~\cite{guo.170930.1}.

 Because of this result from Ref.~\cite{guo.170930.1} we can then properly choose the phase of the
 coupling to a partial wave so that its compositeness is $|X_{\ell S}|$. We then   have
 the following criterion for the elementariness of a narrow resonance with respect to the explicit channel
 considered in the NR treatment [of course, there could be several partial waves and this fact is
 properly taking by the sum over them, cf. Eq.~\eqref{171001.16}]
 \begin{align}
  \label{171001.16}
   |X|\ll 1~.
 \end{align}
 with $X$ calculated as in Eq.~\eqref{171001.7}.
  This criterion cannot be strictly extended
 to a relativistic narrow resonance because the expectation value of an
 operator number $N_D^A$ counts all the particles of type
 $A$ present in any possible open or closed channel.

 The outlined procedure for the  narrow resonance case was generalized in the same reference \cite{guo.170930.1} 
 to a finite width resonance whose pole lies in the Riemann sheet that connects continuously
 with the physical axis between two consecutive channels.
  In NR QFT one should require that
 $E_{{\rm th},n}<M_R<E_{{\rm th},{n+1}}$, while in the relativistic case one should use the $s$ variable and write
 $s_{{\rm th},n}<{\rm Re}s_R<s_{{\rm th},n+1}$, with $s_R=(M_R-i\Gamma/2)^2$, and the thresholds for the channels
 $n$ and $n+1$ are indicated with an obvious notation. 
 The point of this requirement is that the Laurent
 series around the resonance pole can match with  the physical axis within some energy interval,
 so that the modules of the residue at the
 resonance poles have still physical meaning as couplings.
 In order to apply safely this requirement one should
 ascertain a physical process in which the non-resonant terms would play little role and then the resonance
 signal becomes well manifest.
 A good example of this
 is the $f_0(500)$ resonance or $\sigma$ which can hardly be seen in isoscalar scalar $\pi\pi$ scattering
 while it is manifest in the pion scalar form factor which is the one that drives the low-energy
 part of the decays of $D^+\to \pi^-\pi^+\pi^+$ \cite{ref.171001.4}, as discussed in Ref.~\cite{ref.171001.3}.

 We can give another thought (a more ``microscopic'' one)
 for the origin of such phase transformation of the couplings stemming from Eq.~\eqref{171001.15} and
 introduced in Ref.~\cite{guo.170930.1}.
  To accomplish this aim let us  consider 
  energy-dependent transformations in the partial-wave projected in/out states. These are driven 
  by a function $\eta_i(E)$, which at least
  has a unitarity cut and satisfy the Schwarz reflection principle $\eta_i(E\pm i\vep)=\eta_i(E\mp i\vep)^*$,
  as it is the case for partial-wave scattering amplitudes. Here the subscript $i$ refers to any partial
  wave to which the resonance couples. The transformation in question is
 \begin{align}
   \label{171001.16b}
 |\psi_\alpha^+\rangle&\to e^{\eta_i(E_\alpha+i\vep)}|\psi_\alpha^+\rangle~,\\
 \langle \psi_\alpha^-|&\to \langle \psi_\alpha^-| e^{\eta_i(E_\alpha-i\vep)^*}=\langle \psi_\alpha^-|e^{\eta_i(E_\alpha+i\vep)}~.
 \end{align}
 In this way when performing the analytical extrapolation to the 2nd RS to reach the resonance pole
 at $E_R$ we have to cross the unitarity cut and enter in this unphysical sheet so that
 first 
 \begin{align}
   \label{171001.17}
 \eta_i(E_\alpha+i\vep)\to \eta_i^{{II}}(E_\alpha-i\vep)~,
   \end{align}
 and from here, with $E$ having already negative imaginary part, reach $E_R$ with the value
 $\eta^{{II}}(M_R-i\Gamma/2)$. Let us stress that this transformation has no analogue
 for a bound state. As a result of this transformation the couplings change as
 \begin{align}
   \label{171001.18}
   g_i^2(k^2)\to g_i^2(k^2) e^{2\eta_i^{{II}}(E_R)}~.
 \end{align}

 For the case of a narrow resonance we can write a plausible dispersion relation for
 the smooth function $\eta_i(E)$ around the resonance region as
 \begin{align}
   \label{171001.19}
   \eta_i(E)&=\frac{1}{\pi}\int dE'\frac{{\rm Im}\eta_i(E')}{E'-M_R-i\Gamma/2}
   \approx \frac{1}{\pi}  \dashint dE'\frac{{\rm Im}\eta_i(E')}{E'-M_R}
   +i\,{\Im}\eta_i(M_R)~.
 \end{align}
 Since ${\rm Im}\eta_i(E')$ is nearly constant around the narrow-resonance mass, 
 its Cauchy principal value around the latter should be very small
 and the dominant contribution in Eq.~\eqref{171001.19} is its imaginary part.
 Therefore in this case we recover the 
 results of Ref.~\cite{guo.170930.1} and we have the change in the coupling by a phase factor
 \begin{align}
   \label{171001.20}
   g_i^2(k^2)\to g_i^2(k^2) e^{2i{\rm Im}\eta_i^{{II}}(E_R)}~.
 \end{align}
 This derivation also shows that for a finite width resonance is not so clear that $\eta_i(E)$
 is just a purely imaginary number. However, in the lines of the discussion above, for
 a resonance that is manifest on the physical real energy axis the modules of its residues
 can be interpreted as physical
 couplings and the corrections on them (if any) would be relatively small and a transformation like that
 in Eq.~\eqref{171001.20} should be reasonable.

 In summary, for a narrow resonance lying above threshold in NR QFT
 we can calculate its compositeness on an open partial wave by taking the absolute value of $X_{\ell S}$.
 For a finite-width resonance we can say that this is also a reasonable calculation 
 if the resonance is manifest at some interval along the physical energy axis.

 In the relativistic case the situation is a priori
 less clear since one cannot exclude contributions from closed channels containing particles of type $A$ in
 the evaluation of the expectation  value of a number operator $N_D^A$. Therefore,
 the change of phase in the couplings of only the open channels is not of general usage.
 Nonetheless, in practical applications within models that incorporate only a few coupled channels
 and with expected suppression of extra multi-particle components, 
 one could still apply these changes of phase in the couplings for the open channels and give
 physically reasonable results.

 %%%%%%%%%%%%%%%%%%%%%%%%%%%%%%%%%%%%%%%%%%%%%%%%%%%%%%%%%%%%%%%%%%%%%%%%%%%%
\section{Conclusions}
 \label{sec.171001.1}

 We have given a new perspective to the problem of the compositeness/elementariness of
 a bound state or a resonance by considering the expectation values in the state of the number operators
 of the free particle species.
 This new formalism is an important step forward for this relevant problem.

At the fundamental level there are important examples in which the Hamiltonian is not expressed
 in terms of the asymptotic degrees of freedom, e.g. Quantum Chromodynamics (QCD).
  It is then clear that answering the question whether a
  bound or resonance state generated in such theories is elementary or a composite 
 of the asymptotic states in the continuum might be particularly demanding. 
  It could be also the case  that the bare elementary states are integrated out in the effective field theory, 
 so that in relativistic QFT one cannot address then the issue of compositeness of the dynamically-generated
 bound states and resonances  in terms of the traditional language based on the wave-function
 renormalization of the bare elementary  field. 
 Let us stress that one can address both important questions  
 on the composite or elementary nature of a bound or resonance state with respect to the states 
 in the continuum by evaluating the expectation values 
 of the operator numbers of the free particle states in QFT, as developed in this work. 
 These questions correspond indeed to common situations in hadron physics.
    
 We have discussed first the non-relativistic case and developed  suitable expressions
 for its computation within QFT, e.g. by using Feynman diagrams. 
In terms of them we have provided a new closed equation for the compositeness of a non-relativistic bound state in the
 scattering of two particles with large wavelengths compared with the typical range of their interaction. 
This equation has allowed us to conclude that $X=1$ for zero-range energy-independent potentials. 
This conclusion has been also be extended for any finite-range energy-independent potential, being regular or singular. 
 The  equation for the calculation of the expectation values of the number operators
 in a bound state within relativistic QFT can be easily derived from its  NR QFT counterparts.
 In this way a universal criterion for the elementary character of a
 bound state, both in NR and relativistic QFT, has been given for the first time.
 It is also shown that $X$  is independent under unitarity transformations and field redefinitions. 
 We also offered a non-trivial example in the NR QFT case for an energy-dependent potential 
between the asymptotic particles in the continuum in which it is shown that 
$X$ is independent of any type of cutoff regulator employed, once its nonperturbative calculation is undertaken. 
 This case also illustrates the difficulties that dimensional regularization might have in nonperturbative calculations 
(complementing with more examples those already given in Ref.~\cite{phillips.180319.1}).

 Next, we have moved on to the resonance case.  The equations for the calculation of the
 expectation values of the number operators in a resonance state have been given within NR and
 relativistic QFT. We have  deduced as well a universal necessary condition for a resonance  being
 qualified as elementary. Within non-relativistic scattering theory
  we have derived  that $X=1$ for finite-range energy-independent potentials, similarly as for bound states.
  We have also introduced suitable phase-factor transformation
  that are closely related to the $S$-matrix transformations first given in Ref.~\cite{guo.170930.1}.
  In terms of them one can end with real positive values for the compositeness of a narrow resonance
  in NR QFT with respect the open channels.
  This result can be also extended with quite confidence to the case of finite-width resonances
  following the same methods.
  For the relativistic case, the use of unitary transformations only upon the couplings to the open channels 
  is not enough to derive meaningful positive real values of these expectation values in the general case. 
Nonetheless, one should stress that  in many practical examples (e.g. when just a few coupled-channels are
 included in the model) they are  of interest.

\subsection*{Acknowledgments}
 I would like to thank  discussions with Feng-Kun Guo. 
This work is supported in part by the MINECO (Spain) and EU grant FPA2016-77313-P.

\medskip

%%%%%%%%%%%%%%%%%%%%%%%%%%%%%%%%%%%%%%%%%%%%%%%%%%
\appendix

%%%%%%%%%%%%%%%%%%%%%%%%%%%%%%%%%%%%%%%%%%%%%%%%%%%%%%%%%%%%
\section{Decomposition in partial waves in the $\ell S$ basis}
\label{app.170928.1}
\def\theequation{\Alph{section}.\arabic{equation}}
\setcounter{equation}{0}

Let us follow the steps indicated in the footnote \ref{footnote.180603.1} to calculate the diagrams in 
 Fig.~\ref{fig.170930.1}. In QFT we have that their sum is 
\begin{align}
\label{180603.5}
M&= i\int \frac{d^4k}{(2\pi)^4}f(\vk)^2\left\{\frac{(2M_A)^2(2M_B)}{[(P-K)^2-M_A^2]^2(k^2-M_B^2)}
+\frac{(2M_A)(2M_B)^2}{(k^2-M_A^2)[(P-k)^2-M_B^2]}
\right\}~,
\end{align}
where $f(\vk)^2$ is the coupling squared, that only depends on the
 three-momentum in the non-relativistic case,  $\mathbf{P}=0$ and $P^0=M_A+M_B+E_B$ (of course, the masses 
of the particles $A$ and $B$, $M_A$ and $M_B$, in order, are much 
bigger than $|E_B|$). We have also included the right mass factors multiplying the relativistic propagators 
so as to end with the standard NR reduction, corresponding to the standard normalization to 
$(2\pi)^3\delta(\vp-\vq)$ for NR plane-wave states. 
We next perform the integration over $k^0$ in Eq.~\eqref{180603.5}, e.g. by closing the $k^0$
 integration contour along the upper half plane. Let us consider the first term on the r.h.s. 
of the previous equation and introduce the notation $w_i=\sqrt{M_i^2+\vk^2}$. We have
\begin{align}
\label{180603.6}
I_1&=\int\frac{dk^0}{2\pi}\frac{(2M_A)^2(2M_B)}{[(P^0-k^0)^2-w_A^2+i\vep]^2({k^0}^2-w_B^2+i\vep)}~.
\end{align}
This integral has two poles for $\Im k^0>0$, a double one at $k^0=P^0-w_A+i\vep$ and a 
simple pole at $k^0=-w_B+i\vep$. Calculating the residues of these poles we have
\begin{align}
\label{180603.7}
I_1&=i(2M_A)^2(2M_B)\left[\frac{\partial}{\partial k^0}\left\{
\frac{1}{(k^0-P^0-w_A)^2(k^0-w_B+i\vep)(k^0+w_B)}\right\}_{k^0=P^0-w_A+i\vep}\right.\\
&\left.-i\frac{1}{(w_A+w_B+P^0)^2(w_A-w_B-P^0)^2 2 w_B}\right]~.\nn
\end{align}
We now proceed with the NR reduction of every factor in the denominators
\begin{align}
\label{180603.8}
&k^0=P^0-w_A+i\vep\to M_B+\frac{\varkappa^2}{2\mu}-\frac{\vk^2}{2M_A}+i\vep~,\\
&k^0-P^0-w_A\to -2 M_A~,\nn\\
&k^0-w_B+i\vep\to \frac{\varkappa^2}{2\mu}-\frac{\vk^2}{2\mu}+i\vep~,\nn\\
&k^0+w_B \to 2M_B~,\nn\\
&w_A+w_B+P^0\to 2(M_A+M_B)~,\nn\\
&w_A-w_B-P^0\to -2M_B~,\nn\\
&2w_B\to 2M_B~. 
\end{align}
As a result
\begin{align}
\label{180603.9}
I_1\to \frac{-i}{(\varkappa^2/2\mu-\vk^2/2\mu+i\vep)^2}~.
\end{align}
For the second term on the r.h.s. of Eq.~\eqref{180603.5} we have the same result. 
Summing both contributions we obtain 
\begin{align}
\label{180603.10}
X&=\int \frac{d^3\vk}{(2\pi)^3}\frac{f(\vk)^2}{(\varkappa^2/2\mu-\vk^2/2\mu+i\vep)^2}~.
\end{align}
For a bound state the $+i\vep$ can be dropped because $\varkappa^2<0$.

The coupling of a resonance with angular momentum $J$ to a two-particle channel with
orbital angular momentum $\ell$ and total spin $S$ is
\begin{align}
  \sqrt{4\pi}\,g_{\ell S}\sum_{m,M}(\sigma_1\sigma_2M|s_1s_2S)(mM\mu|\ell S J)Y_\ell^m(\hat{\vp})
\end{align}
When using this decomposition of the coupling into the equation for calculating $X$, which
depends on the coupling squared,  the diagonal sum over $X_{\ell S}$ in Eq.~\eqref{170930.5} results 
once the angular integration and the sum over the $\sigma_i$, $i=1,2$, are performed. 
 Here one has to use the orthogonality properties of the
 spherical harmonics and the Clebsch-Gordan coefficients.

%%%%%%%%%%%%%%%%%%%%%%%%%%%%%%%%%%%%%%%%%%%%%%%%%%%%%%%%%%%%%%%%%%%%%%%%%%%%%%%%%%%%%%%%%%%%

\end{document}